\documentclass[twocolumn]{aastex631}

\usepackage{epsfig,graphics,subfigure,psfrag,amsmath,amssymb,amscd,ulem}
\usepackage{appendix}

\begin{document}

\title{Unveiling the Cosmic Dance of Repeated Nuclear Transient ASASSN-14ko: Insights from Multiwavelength Observations}

\author[0000-0001-7689-6382]{Shifeng Huang}
\affiliation{CAS Key Laboratory for Research in Galaxies and Cosmology, Department of Astronomy, University of Science and Technology of China, Hefei, 230026, China; sfhuang999@ustc.edu.cn, jnac@ustc.edu.cn,twang@ustc.edu.cn}
\affiliation{School of Astronomy and Space Sciences,
University of Science and Technology of China, Hefei, 230026, China}

\author[0000-0002-1517-6792]{Tinggui Wang}
\affiliation{CAS Key Laboratory for Research in Galaxies and Cosmology, Department of Astronomy, University of Science and Technology of China, Hefei, 230026, China; sfhuang999@ustc.edu.cn, jnac@ustc.edu.cn,twang@ustc.edu.cn}
\affiliation{School of Astronomy and Space Sciences,
University of Science and Technology of China, Hefei, 230026, China}

\author[0000-0002-7152-3621]{Ning Jiang}
\affiliation{CAS Key Laboratory for Research in Galaxies and Cosmology, Department of Astronomy, University of Science and Technology of China, Hefei, 230026, China; sfhuang999@ustc.edu.cn, jnac@ustc.edu.cn,twang@ustc.edu.cn}
\affiliation{School of Astronomy and Space Sciences,
University of Science and Technology of China, Hefei, 230026, China}

\author[0000-0001-5012-2362]{Rong-Feng Shen}
\affiliation{School of Physics and Astronomy, Sun Yat-sen University, Zhuhai 519082, People's Republic of China}
\affiliation{CSST Science Center for the Guangdong-Hong Kong-Macau Greater Bay Area, Sun Yat-sen University, Zhuhai 519082, People's Republic of China}

\author[0009-0006-5497-8049]{Zhaohao Chen}
\affiliation{Shandong Provincial Key Laboratory of Optical Astronomy and Solar-Terrestrial Environment, School of Space Science and Physics, Institute of Space Sciences, Shandong University, Weihai 264209, People's Republic of China}

\author[0000-0003-0203-1196]{Yuanming Wang}
\affiliation{Centre for Astrophysics and Supercomputing, Swinburne University of Technology, Hawthorn, Victoria, 3122, Australia}
\affiliation{ARC Centre of Excellence for Gravitational Wave Discovery (OzGrav), Hawthorn, Victoria, 3122, Australia}

\author[0000-0003-3824-9496]{Jiazheng Zhu}
\affiliation{CAS Key Laboratory for Research in Galaxies and Cosmology, Department of Astronomy, University of Science and Technology of China, Hefei, 230026, China; sfhuang999@ustc.edu.cn, jnac@ustc.edu.cn,twang@ustc.edu.cn}
\affiliation{School of Astronomy and Space Sciences,
University of Science and Technology of China, Hefei, 230026, China}

\author[0000-0003-4225-5442]{Yibo~Wang}
\affiliation{CAS Key Laboratory for Research in Galaxies and Cosmology, Department of Astronomy, University of Science and Technology of China, Hefei, 230026, China; sfhuang999@ustc.edu.cn, jnac@ustc.edu.cn,twang@ustc.edu.cn}
\affiliation{School of Astronomy and Space Sciences,
University of Science and Technology of China, Hefei, 230026, China}

\author[0000-0003-2679-0445]{Yunguo Jiang}
\affiliation{Shandong Provincial Key Laboratory of Optical Astronomy and Solar-Terrestrial Environment, School of Space Science and Physics, Institute of Space Sciences, Shandong University, Weihai 264209, People's Republic of China}

\author[0000-0002-7020-4290]{Xinwen Shu}
\affiliation{Department of Physics, Anhui Normal University, Wuhu, Anhui, 241002, People's Republic of China}

\author[0009-0001-8733-2088]{Hucheng Ding}
\affiliation{Department of Physics, Anhui Normal University, Wuhu, Anhui, 241002, People's Republic of China}

\author[0000-0002-4270-3332]{Xiongjun Fang}
\affiliation{Department of Physics, Key Laboratory of Low Dimensional Quantum Structures and Quantum Control of Ministry of Education, and Synergetic Innovation Center for Quantum Effects and Applications, Institute of Interdisciplinary Studies, Hunan Normal University, Changsha 410081, China}

\author[0000-0001-7142-7667]{Yifan Wang}
\affiliation{College of Physics and Electronic Information, Dezhou University, Dezhou 253023, China}

\author[0000-0003-3965-6931]{Jie Lin}
\affiliation{CAS Key Laboratory for Research in Galaxies and Cosmology, Department of Astronomy, University of Science and Technology of China, Hefei, 230026, China; sfhuang999@ustc.edu.cn, jnac@ustc.edu.cn,twang@ustc.edu.cn}
\affiliation{School of Astronomy and Space Sciences,
University of Science and Technology of China, Hefei, 230026, China}

\author[0000-0003-2477-3430]{Jingran Xu}
\affiliation{Shandong Provincial Key Laboratory of Optical Astronomy and Solar-Terrestrial Environment, School of Space Science and Physics, Institute of Space Sciences, Shandong University, Weihai 264209, People's Republic of China}

\author[0000-0001-5603-7521]{Xu~Chen}
\affiliation{Shandong Provincial Key Laboratory of Optical Astronomy and Solar-Terrestrial Environment, School of Space Science and Physics, Institute of Space Sciences, Shandong University, Weihai 264209, People's Republic of China}

\author[0000-0003-4959-1625]{Zheyu Lin}
\affiliation{CAS Key Laboratory for Research in Galaxies and Cosmology, Department of Astronomy, University of Science and Technology of China, Hefei, 230026, China; sfhuang999@ustc.edu.cn, jnac@ustc.edu.cn,twang@ustc.edu.cn}
\affiliation{School of Astronomy and Space Sciences,
University of Science and Technology of China, Hefei, 230026, China}

\author[0000-0001-6938-8670]{Zhenfeng~Sheng}
\affiliation{Institute of Deep Space Sciences, Deep Space Exploration Laboratory, Hefei, 230026, China}

\begin{abstract}
ASASSN-14ko is a periodically repeating nuclear transient. We conducted high-cadence, multiwavelength observations of this source, revealing several recurrent early bumps and rebrightenings in its UV/optical light curves. The energy released during these bumps and rebrightenings shows a diminishing trend in recent UV/optical outbursts, which we monitored through multiwavelength observations. These features can be ascribed to the interaction between stream debris and the expanded disk in the repeated partial tidal disruption event. The X-ray light curve exhibits an inverse pattern compared to the UV/optical bands, displaying sporadic outbursts. Furthermore, our observations demonstrate that the blackbody temperature and radius in each outburst increase with the UV/optical luminosity, and such evolution resembles that observed in X-ray quasiperiodic eruptions, whereas distinguishing it from typical tidal disruption events.

\end{abstract}

\keywords{Tidal disruption (1696) --- Supermassive black holes (1663) --- Black hole physics (159) --- Accretion (14)}

\section{Introduction} \label{sec:intro}
A tidal disruption event (TDE) occurs when a star wanders too close to a black hole (BH). To date, more than 100 TDEs or TDE candidates have been identified by a variety of time-domain surveys\citep{Velzen2019,Velzen2021,Gezari2021,Jiang2021,Wang2022,Hammerstein2023,Yao2023,Zhu2023,Huang2024}. The majority of these events were discovered in optical bands, characterized by a ``fast rise and slow decay" pattern in their light curves \citep{Yao2023,Hammerstein2023}, along with or without broad emission lines (e.g. H$\alpha$ and He \textsc{ii}) and a blue continuum with a nearly constant and high blackbody temperature \citep{Arcavi2014,Velzen2020,Gezari2021}.  

High-cadence monitoring has revealed that a subset of TDEs exhibit an early ``bump" or precursor flare in their UV/optical light curves, as observed in events such as ASASSN-19bt \citep{Holoien2019}, AT~2019azh \citep{Faris2023}, ASASSN-18ap \citep{Wang2024}, AT~2020wey \citep{Charalampopoulos2023} and especially, AT~2023lli \citep{Huang2024}. Furthermore, \cite{Wang2024} presents a collection of TDEs and candidates featuring a ``bump" during the rising phase of their UV/optical light curves. Among these, AT~2023lli displays the most pronounced early bump, lasting one month and occurring approximately two months prior to the main outburst. \cite{Huang2024} suggested that such a distinct early bump could stem from stream-stream interactions or a double TDE scenario. Several recent theoretical models, such as stream-wind interactions \citep{Calderon2024} and the ``TDE encore" \citep{Ryu2024}, have successfully replicated these early bumps. In addition, mechanisms such as the cooling of unbound debris, vertical compression during the initial passage, and shock breakout from self-intersecting debris streams have been proposed to account for these phenomena \citep{Kasen2010,Yalinewich2019,Wang2024}.

In recent years, a growing number of rebrightening events in TDE light curves have been detected \citep{Jiang2019,Yao2023}. The circularization of debris and the delayed accretion after the disruption are posited as potential explanations for these occurrences \citep{Chen2022,Wang2023}. Notably, events such as AT~2018fyk \citep{Wevers2023,Pasham2024}, eRASSt~J045650.3-203750 \citep{Liu2023b,Liu2024}, RX~J133157.6-324319.7 \citep{Malyali2023}, ASASSN-14ko \citep{Payne2021,Payne2022,Payne2023,Tucker2021}, IRAS~F01004-2237 \citep{Sun2024}, AT~2020vdq \citep{Somalwar2023,Bandopadhyay2024}, AT~2022dbl \citep{Lin2024,Hinkle2024,Zhong2024}, AT~2019aalc \citep{Veres2024}, AT~2021aeuk \citep{Bao2024,Sun2025} and AT~2023adr \citep{Llamas2024} have exhibited a secondary outburst occurring hundreds to thousands of days after the initial flare. These events are considered to be possible repeated partial TDEs (rpTDEs). Intriguingly, GSN~069 also displays signatures of rpTDE \citep{Miniutti2023a,Sheng2021}, potentially linked to the quasi-periodic eruptions (QPEs) observed in this source \citep{Miniutti2019}. 

ASASSN-14ko was first identified as a nuclear transient in the Seyfert galaxy ESO~253-G003 \citep{Holoien2014}, with a redshift of 0.042 and the central black hole mass of $10^{7.85}\, M_\odot$ \citep{Payne2021}, exhibiting a remarkable periodicity of about 115 days, accompanied by a negative period derivative of -0.0026 existing in its UV/optical bands \citep{Payne2021,Payne2022,Payne2023}. Several theoretical studies have explored the mechanisms behind this periodicity, such as the rpTDE process \citep{Liu2023,LiuC2025,Bandopadhyay2024} and star-disk interactions \citep{Linial2024}. In the context of rpTDEs, \cite{Cufari2022} suggested that the disrupted star originated from a binary system via the Hill mechanism. Analogously to rpTDEs, periodic mass transfers from a star or binary system to a BH have been proposed as potential sources of similar signals \citep{Shen2019,King2023,Metzger2022,Chen2023}. Initially, its UV/optical light curves were characterized by a smooth ``fast rise and slow decay" pattern. However, subsequent high-cadence multiwavelength observations have revealed intricate details within these light curves. \cite{Huang2023} observed repeated early bumps during the rise and subsequent rebrightenings in the decline phase of two separate outbursts. These particular phenomena are likely the result of interactions between stream debris and an expanded accretion disk during rpTDE events. 

In this work, we present comprehensive multiwavelength observations of ASASSN-14ko, spanning from radio to X-rays. We explore the dynamics of recurrent early bumps and subsequent rebrightenings. Our high-cadence observations reveal notable quasiperiodic patterns in the X-ray light curve, which appear to correlate with the UV/optical bands. In Section \ref{sec:data}, we detail our data reduction process, followed by a presentation of the new observational results in Section \ref{sec:results}. Section \ref{sec:discussion} is dedicated to discussing the underlying physical mechanisms of ASASSN-14ko, focusing on the emergence of early bumps and rebrightenings, the quasi-periodicity observed in the X-ray domain, and the dynamic characteristics of the system. In Section \ref{sec:conclusion}, we present a summary of the results and conclusions. For this work, we adopt the cosmological parameters of $H_0=70\,\text{km}\text{s}^{-1},\text{Mpc}^{-1}$, $\Omega_{\rm M}=0.3$, and $\Omega_{\Lambda}=0.7$.

\section{Observations and Data Reduction} \label{sec:data}
\subsection{Swift X-Ray data}
The X-Ray Telescope (XRT), which has a sensitivity range of 0.3 to 10.0 keV, onboard the Neil Gehrels \emph{Swift} Observatory was utilized to monitor the flux and spectral evolution of ASASSN-14ko in the X-ray band\citep{Burrows2005}. Since 2022 December 2, we requested 172 ToO observations, resulting in a cadence of 2-5 days. The public data (PIs: Payne and Chakraborty) were also retrieved from the High Energy Astrophysics Science Archive Research Center (HEASARC) website and analyzed. The \emph{Swift} data were processed with \texttt{HEASoft 6.33.1}. We ran the task \texttt{xrtpipeline} and \texttt{xrtproducts}, respectively, to generate the light curves and spectra. A circular region with a radius of $47.1^{\prime\prime}$ was used to extract the source region centered on the object, while the background region was extracted using an adjacent circular area with a radius of $120^{\prime\prime}$.

\subsection{Chandra data reduction}
We retrieved the high spatial resolution archival data from Chandra on August 18, 2023, and September 5, 2023 (\dataset[ObsID 26384 and 27453, PI: Schartel]{https://doi.org/10.25574/cdc.391}. The task \texttt{chandra\_repro} in CIAO 4.15 was executed to reprocess the data, followed by running \texttt{specextract} to derive the X-ray spectra. A weak active galactic nucleus lies adjacent to ASASSN-14ko \citep{Tucker2021}. Although it remains unresolved in Swift images, it is clearly visible in Chandra images \citep{Payne2023}. The source regions for both ASASSN-14ko and the nearby object were extracted using circles with a radius of $1.5^{\prime\prime}$, while the background region was extracted using a circle with a radius of $30^{\prime\prime}$.

\subsection{Swift UV/Optical Photometry}
The UV/Optical Telescope (UVOT) onboard the \emph{Swift} observatory is equipped with seven filters \citep[\textsl{V}, \textsl{B}, \textsl{U}, \textsl{UVW1}, \textsl{UVM2}, \textsl{UVW2} and white;][]{Roming2005}. We ran \texttt{uvotimsum} to stack the images and then, with the source and background regions defined by circles with radii of $10^{\prime\prime}$ and $40^{\prime\prime}$, we executed the task \texttt{uvotsource} to generate the light curves. Based on \cite{Schlafly2011} and utilizing the online tool\footnote{\url{https://irsa.ipac.caltech.edu/applications/DUST/}}, we determined a Galactic extinction value of $E(B-V)=0.043$. Subsequently, we adjusted the magnitudes for each band according to the extinction law of \cite{Cardelli1989}. In this study, we adopt Galactic extinction values of 0.14, 0.18, 0.21, 0.27, 0.42, and 0.41 mag for the \textsl{V}, \textsl{B}, \textsl{U}, \textsl{UVW1}, \textsl{UVM2}, and \textsl{UVW2} bands, respectively.

\subsection{Ground-based Optical Photometry}
The public g band data points were downloaded from the sky patrol system of All-Sky Automated Survey for Supernovae (ASASSN)\footnote{\url{https://asas-sn.osu.edu/}}. Image subtraction photometry was selected and no reference flux was added to our data \citep{Shappee2014,Kochanek2017}.

We collected host-subtracted light curves of ASASSN-14ko from the Asteroid Terrestrial Impact Last Alert System (ATLAS; \citealt{ATLAS}).
The ATLAS $c$ and $o$ band light curves were obtained using the ATLAS Forced Photometry Service, which produces PSF photometry on the difference images. ATLAS has $3-4$ single exposures within each epoch (typically within one day), so we binned the light curve every epoch to improve the signal-to-noise ratio (SNR).

\subsection{Radio Observations}

We noted that ASASSN-14ko has been observed in multiple epochs by the ASKAP Variables and Slow Transients (VAST; \citealt{Murphy2021PASA...38...54M}) survey and the Rapid ASKAP Continuum Survey (RACS; \citealt{McConnell2020PASA...37...48M,Hale2021PASA...38...58H}) since 2020. 
Each observation had an integration time of approximately 12 min, resulting in a typical rms sensitivity of 0.25\,mJy\,beam$^{-1}$ and an angular resolution of $12''$ at a central frequency of 887.5\,MHz. 
We obtained the flux density and local rms from their catalogs, downloaded from the CSIRO ASKAP Science Data Archive (CASDA\footnote{\url{https://research.csiro.au/casda/}}). 
The catalogues were extracted from ASKAP mosaiced images, with \texttt{Selavy} \citep{Whiting2012PASA...29..371W} used for source finding and flux measurement.

\section{Results}\label{sec:results}
\subsection{Multiwavelength Light Curves}
\begin{figure*}
    \centering
    \includegraphics[width=0.9\textwidth]{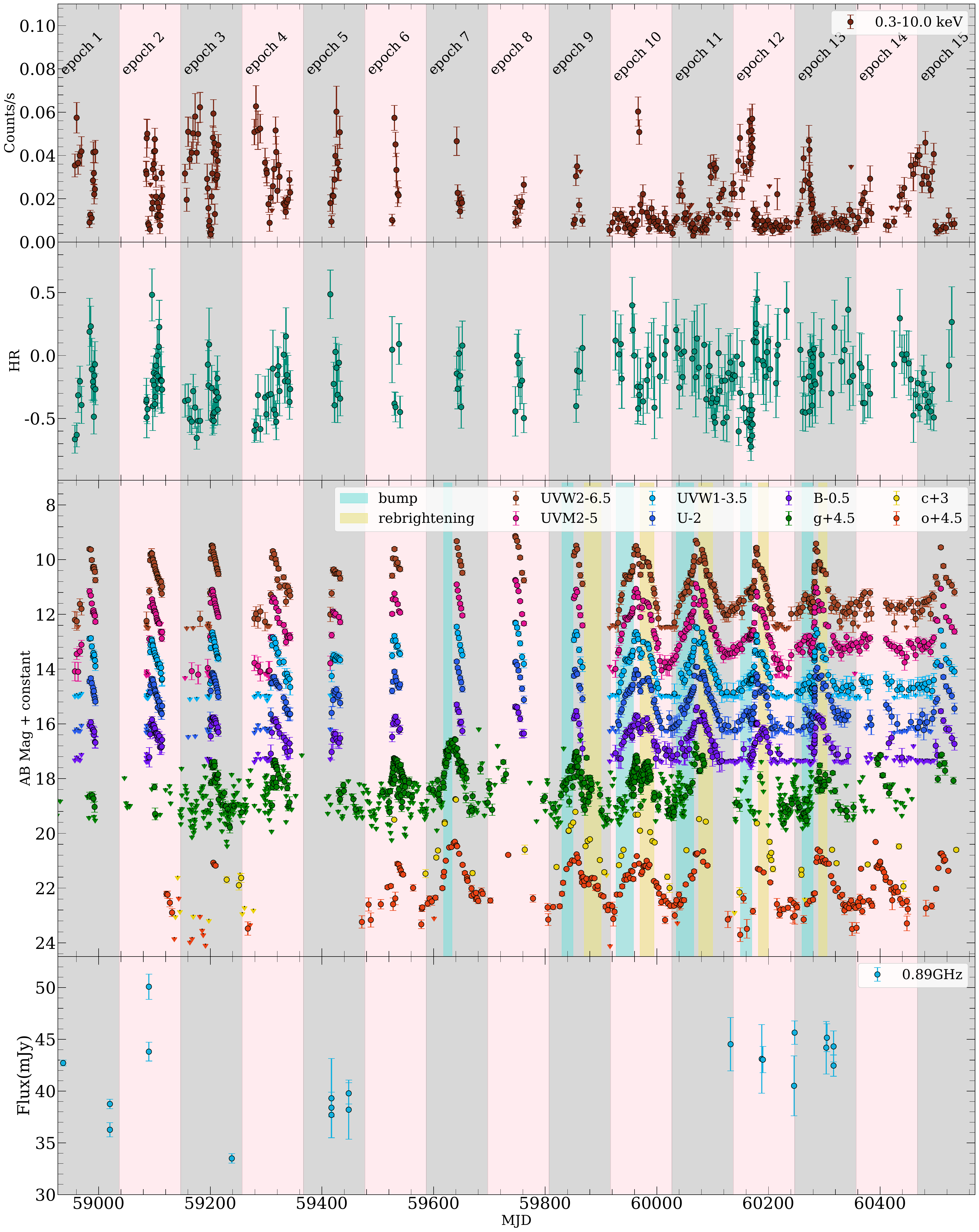}
    \caption{Multiwavelength light curves of ASASSN-14ko. The X-ray light curve, hardness ratio, UV/optical and radio light curves are plotted from the top to the bottom panel. All UV/optical light curves are host-subtracted. Triangles represent the 3$\sigma$ upper limits. Although Chandra can resolve the nearby source, Swift/XRT cannot. Therefore, we present only the observed X-ray light curve and hardness ratio without subtracting the contribution from the nearby source.}
    \label{fig:lc_mw}
\end{figure*}

Since November 2022, we performed high-cadence multiwavelength observations for ASASSN-14ko through Swift and multiple outbursts were monitored. Surprisingly, repeated bumps and rebrightenings in the rising and decreasing UV/optical light curves were also detected in the last several outbursts \citep{Huang2023}. For convenience, the period MJD 58927--60540 was divided into 15 epochs. We carefully checked the historical data of ASASSN and ATLAS, the specific structure with bump had been recorded in epochs 7 and 9. The enriched multiwavelength monitoring for this source showed a smooth ``fast rise and slow decay" tendency before these epochs \citep{Payne2021,Payne2022,Payne2023}. However, in recent outbursts, the UV/optical light curves show that the bumps and rebrightenings have progressively weakened, approaching previous cases. The multiwavelength light curve of ASASSN-14ko can be seen in Figure~\ref{fig:lc_mw}. 

High-cadence X-ray monitoring provides detailed information about the variability. We observed pronounced variations, and X-ray outbursts had shorter durations compared to those in the UV/optical bands. During epochs 10--15, the peak-to-peak intervals in the X-rays exhibited significant deviations from those observed in the UV/optical bands. Across epochs 10--15, the hardness ratio (HR) \footnote{The hardness ratio is defined by (H-S)/(H+S), where H and S denote the count rate of the hard X-ray in 2.0--10.0 keV and the soft X-ray in 0.3--2.0 keV, respectively.} displayed a ``softer-when-brighter" trend, consistent with the findings reported by \cite{Payne2023}.

The source is also detectable in the radio band. During epochs 1--3, ASKAP observed significant variability, reaching a peak flux of $50.09\pm{1.22}~\rm mJy$ on MJD 59089.95, followed by a decline to $33.5\pm{0.45}~\rm mJy$ on MJD 59238.53. Between MJD 60132 and 60316.6, the radio flux remains relatively stable, with a mean value of $43.66\pm{6.36}~\rm mJy$. Unfortunately, the sparse radio monitoring limits our ability to obtain detailed evolutionary information about this source.

\subsection{X-ray spectral analysis}
The high-cadence observations obtained by \emph{Swift} provide an excellent opportunity for the analysis of spectral behavior in high- and low-state X-rays. We define the high state as a net count rate exceeding 0.01 cts/s as measured by \emph{Swift/XRT}, while values below this threshold are considered to represent the low state. 

The two Chandra observations were conducted during the low state. For each observation ID, we extracted the Chandra spectra at the positions of ASASSN‑14ko and the nearby source, then combined the two spectra for each source before performing spectral fitting. We adopted the \texttt{tbabs*zashift*zxipcf*(powerlaw+gaussian)} model to fit the nearby source spectrum, fixing the Galactic column density at $3.49\times 10^{20}\,\text{cm}^{-2}$ \citep{HI4PI2016}. The column density of the ionized absorption material is $6.46_{-4.65}^{+6.71}\times 10^{23}\,\text{cm}^{-2}$, the ionization parameter is $\log \xi=1.91_{-0.87}^{+0.08}\,\text{erg}\,\text{cm}\,\text{s}^{-1}$, the covering fraction is $0.98\pm{0.02}$, power-law photon index of $0.61_{-0.67}^{+0.98}$ with $\chi^2/\text{dof}=32.57/43$. During this fitting, we also determined that the emission line energy in the hard band is $6.34\pm{0.07}\,\text{keV}$ with a width of $0.12^{+0.06}_{-0.07}\,\text{keV}$. Because of the low SNR of the spectra, the errors in the fitted parameters are relatively large, and we are unable to constrain the properties of the absorption materials accurately.
We also fitted the Chandra spectrum of the center of ASASSN-14ko using the \texttt{tbabs*zashift*powerlaw} model, again fixing the Galactic column density at $3.49\times 10^{20}\,\text{cm}^{-2}$. The derived power-law photon index is $2.08_{-0.33}^{+0.34}$, with a C-stat/dof=19.57/20. 

\begin{figure}
    \centering
    \includegraphics[width=0.45\textwidth]{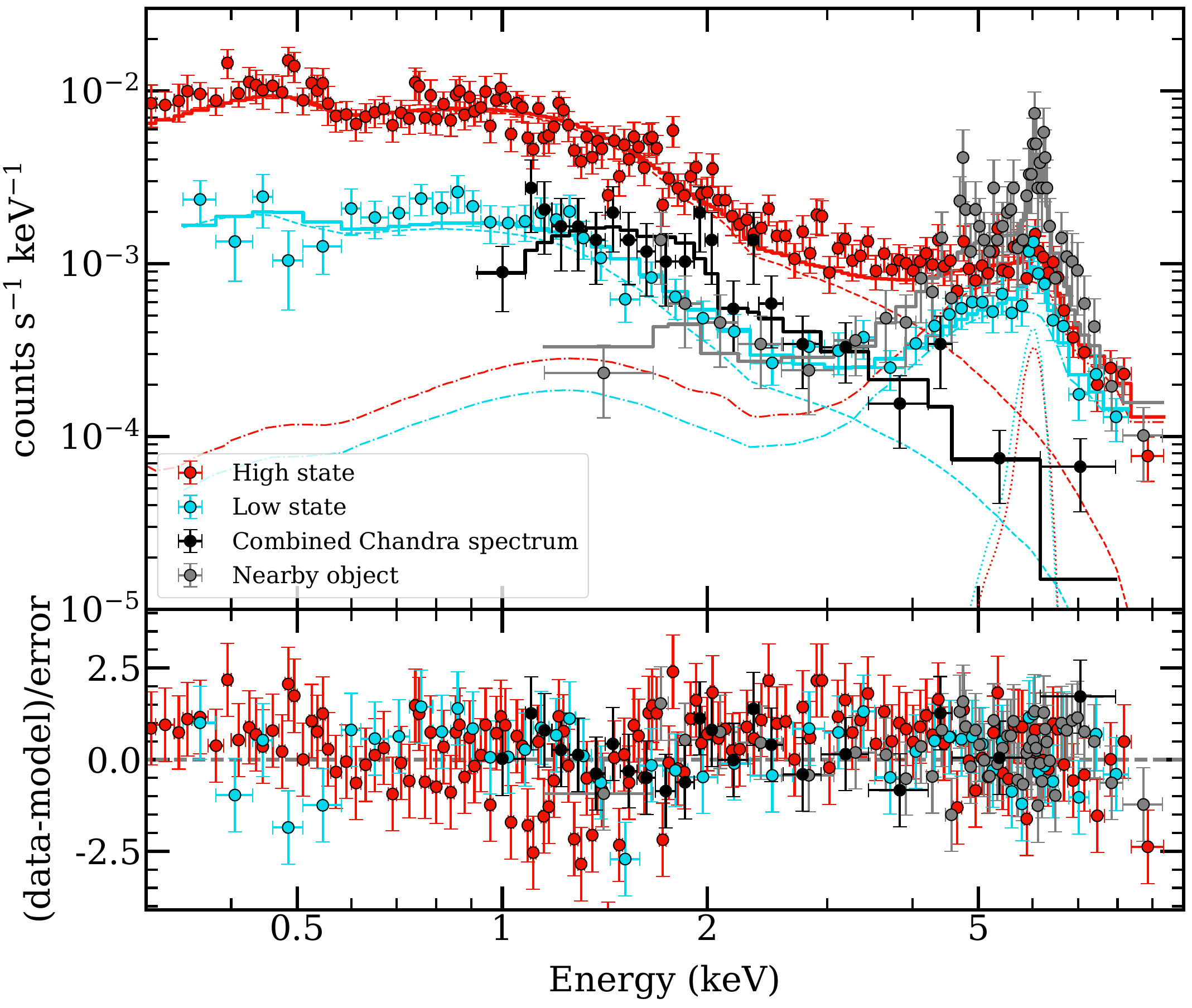}
    \caption{The spectral fitting results for the X-ray spectra. The Swift/XRT spectra were obtained by the high and low states, respectively. The Chandra spectra were derived in the low state for both the central of the nuclear transient and the nearby object. The XRT spectra were fitted with
\texttt{tbabs*zashift*powerlaw + tbabs*zashift*zxipcf*(powerlaw+gaussian)},
the combined Chandra spectrum of ASASSN‑14ko with
\texttt{tbabs*zashift*powerlaw},
and the Chandra spectrum of the nearby source with
\texttt{tbabs*zashift*zxipcf*(powerlaw+gaussian)}}
    \label{fig:xspec}
\end{figure}

Event files from MJD 59910 to 60533 were stacked separately for the high and low states to extract their respective X-ray spectra. Both spectra display an Fe K$\alpha$ emission line in the hard band (2.0--10.0 keV). These X-ray spectra are shown in Figure \ref{fig:xspec}.
The nearby source cannot be separated from the XRT image, and thus the Swift spectra include photons from this nearby source. Consequently, the stacked Swift spectra for both the high and low states were fitted using the \texttt{tbabs*zashift*powerlaw+ tbabs*zashift*zxipcf*(powerlaw+gaussian)} model. To minimize the influence of the nearby source, we assumed that its contribution remained roughly constant. Therefore, the primary parameters for the nearby source component in the spectrum were fixed at the values obtained from the Chandra spectral fitting. For the low state, we derived a photon index of $2.11\pm{0.19}$ ($\chi^2/\text{dof}=37.88/45$), and for the high state, we obtained a photon index of $2.08\pm{0.08}$ ($\chi^2/\text{dof}=197.17/143$). Furthermore, we attempted to add a \texttt{diskbb} component in the fitting and derived a temperature of $0.16\pm{0.02}\,\text{keV}$ and a photon index of $1.40\pm{0.18}$, with a $\chi^2/\text{dof}=140.51/141$ for the high state. However, when applying this model to the low state spectrum, we encountered significant residuals. 

Through spectral fitting, we interpret that the detected X-ray variability in the soft band is primarily attributed to the center of ASASSN-14ko. The emission from the nearby source mainly contributes to the hard-band spectral features.  

\begin{figure}
    \centering
    \includegraphics[width=0.45\textwidth]{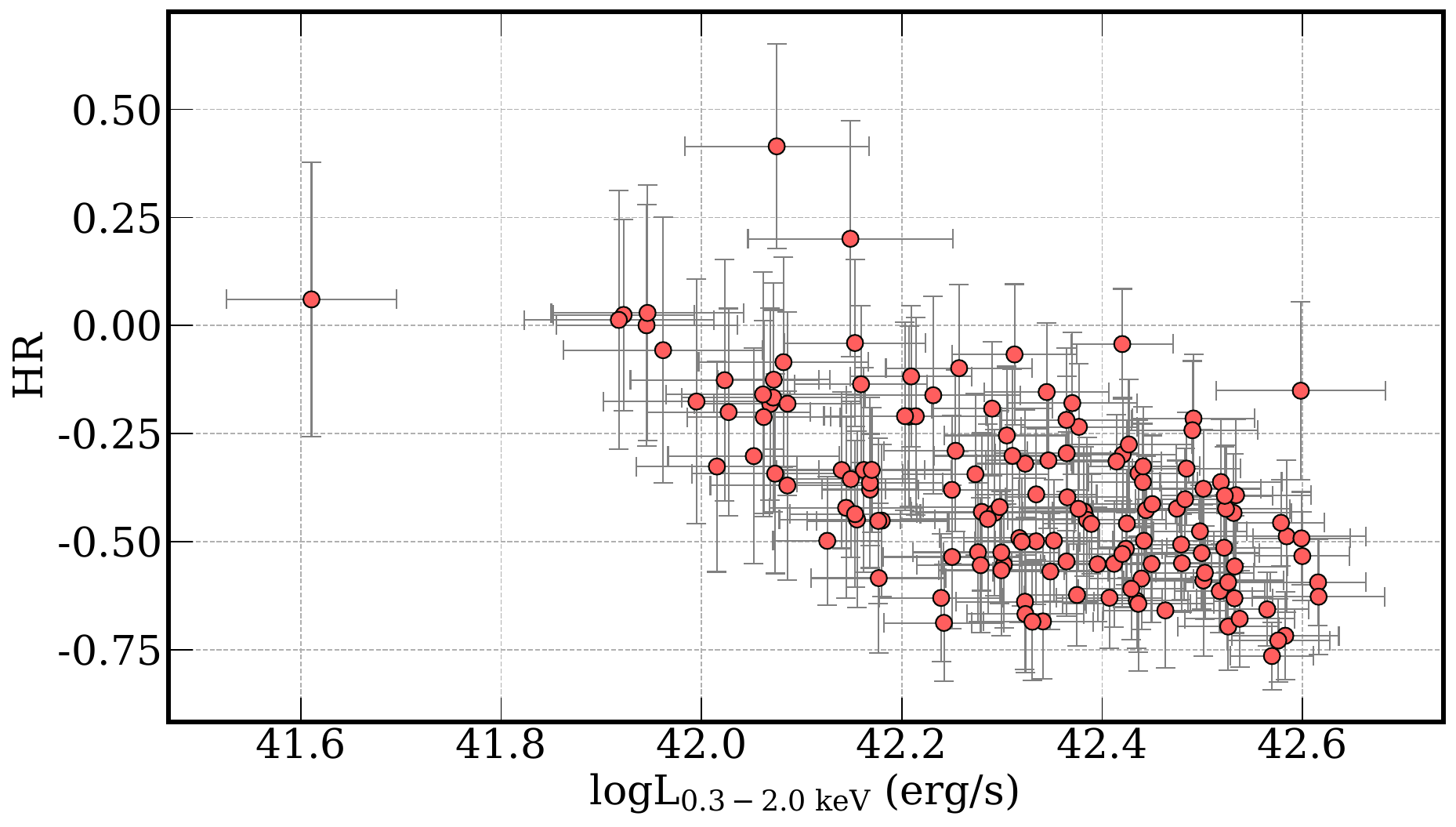}
    \caption{The correlation between HR and luminosity at 0.3--2.0 keV. The contribution from the nearby source has been subtracted. We define the HR by (H-S)/(H+S), in which H denotes the count rate in the hard X‑ray band (2.0--10.0 keV) and S denotes the count rate in the soft X‑ray band (0.3--2.0 keV).}
    \label{fig:HR_L}
\end{figure}

Using the \texttt{xspec} fitting results, we modeled the nearby source and obtained its Swift/XRT count rates: $3.6\times 10^{-4}$ cts/s in the 0.3--2.0 keV band and $3.05\times 10^{-3}$ cts/s in the 2.0--10.0 keV band. After subtracting this contribution, we derived the corrected X-ray light curve and HR, as shown in Figure \ref{fig:HR_L}. The source exhibits a slight ``softer-when-brighter" trend, consistent with the findings of \citet{Payne2023}.

\subsection{Repeated Bumps and Rebrightenings}
\cite{Huang2023} had reported two repeated early bumps and rebrightenings that occurred in epochs 10 and 11. In this work, we find more such variation structures in UV/optical outbursts. An early bump was also observed in the ATLAS historical light curves of the o band during epoch 7, and in the ASASSN g-band. Unfortunately, the Swift UVOT light curve does not cover this time interval. However, UV/optical light curves in epoch 7 declined smoothly and no significant rebrightening was detected. In the UV/optical bands, there were no indications of bumps or rebrightening events detected before epoch 7.

\begin{figure}
    \centering
\includegraphics[width=0.45\textwidth]{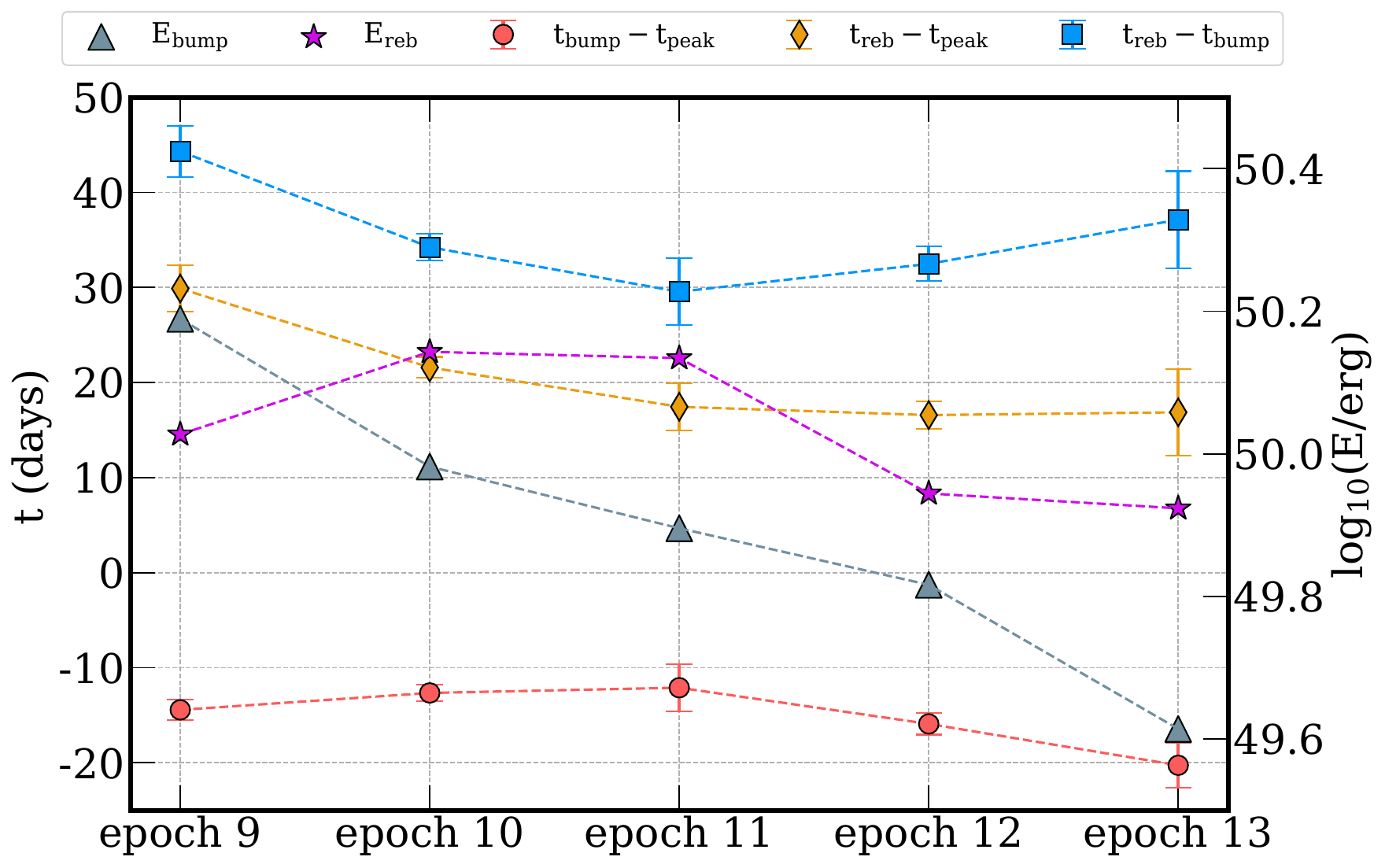}
    \caption{Temporal evolution of energy released in bumps and rebrightenings, occurring time of bumps, rebrightenings and their gap intervals. Here we only consider the epochs with both bumps and rebrightenings.}
\label{fig:intervals}
\end{figure}

The UV/optical photometric data were fitted by the blackbody model through \texttt{SuperBol} \citep{Nicholl2018} to derive the temporal evolution of blackbody luminosity ($L_{\rm bb}$), temperature ($T_{\rm bb}$) and blackbody radius ($R_{\rm bb}$). The results in epoch 7 and epochs 9--13 are shown in the appendix (Figure \ref{fig:lc_bb}). The significant repeated bumps and rebrightenings can be seen in the evolution of $L_{\rm bb}$. $L_{\rm bb}$ shows a series of pronounced bumps and subsequent rebrightenings. These bumps, which occur at times $t_{\rm bump}$, display varying time intervals relative to the UV/optical peak times $t_{\rm peak}$ for each outburst. For instance, during epoch 9, the UV/optical bump manifested 14 days before the peak, whereas in epoch 11, it appeared 12 days before the peak. In a contrasting trend, epoch 13 saw the bump emerge 20 days before the peak, indicating a variable $t_{\rm bump}$ in these events. However, $t_{\rm reb}-t_{\rm peak}$ shows an overall shorter trend. Additionally, the time interval between bumps and rebrightenings $t_{\rm reb}-t_{\rm bump}$, exhibits a tendency to decrease and then increase. Here the $t_{\rm bump}$ and $t_{\rm reb}$ were defined as the central time when the ``bumps" and ``rebrightenings" light curves were fitted by Gaussian function. We derived the energy of bumps ($E_{\rm bump}$) and rebightenings ($E_{\rm reb}$) respectively by integrating the blackbody luminosity. We found that $E_{\rm bump}$ shows a trend of decrease, while the value of $1.55\times10^{50}~\rm erg$ in epoch 9 and $4.11\times10^{49}~\rm erg$ in epoch 13. Unlike bumps, $E_{\rm reb}$ initially increases from $1.06\times10^{50}~\rm erg$ (epoch 9) to $1.39\times10^{50}~\rm erg$ (epoch 10), and then decreases to a value of $8.39\times10^{49}~\rm erg$ in epoch 13. More details can be seen in Figure \ref{fig:intervals}. 

\subsection{Temporal Evolution of Blackbody Temperature and Radius}
\begin{figure*}
    \centering
    \subfigure[]{
    \includegraphics[width=0.75\textwidth]{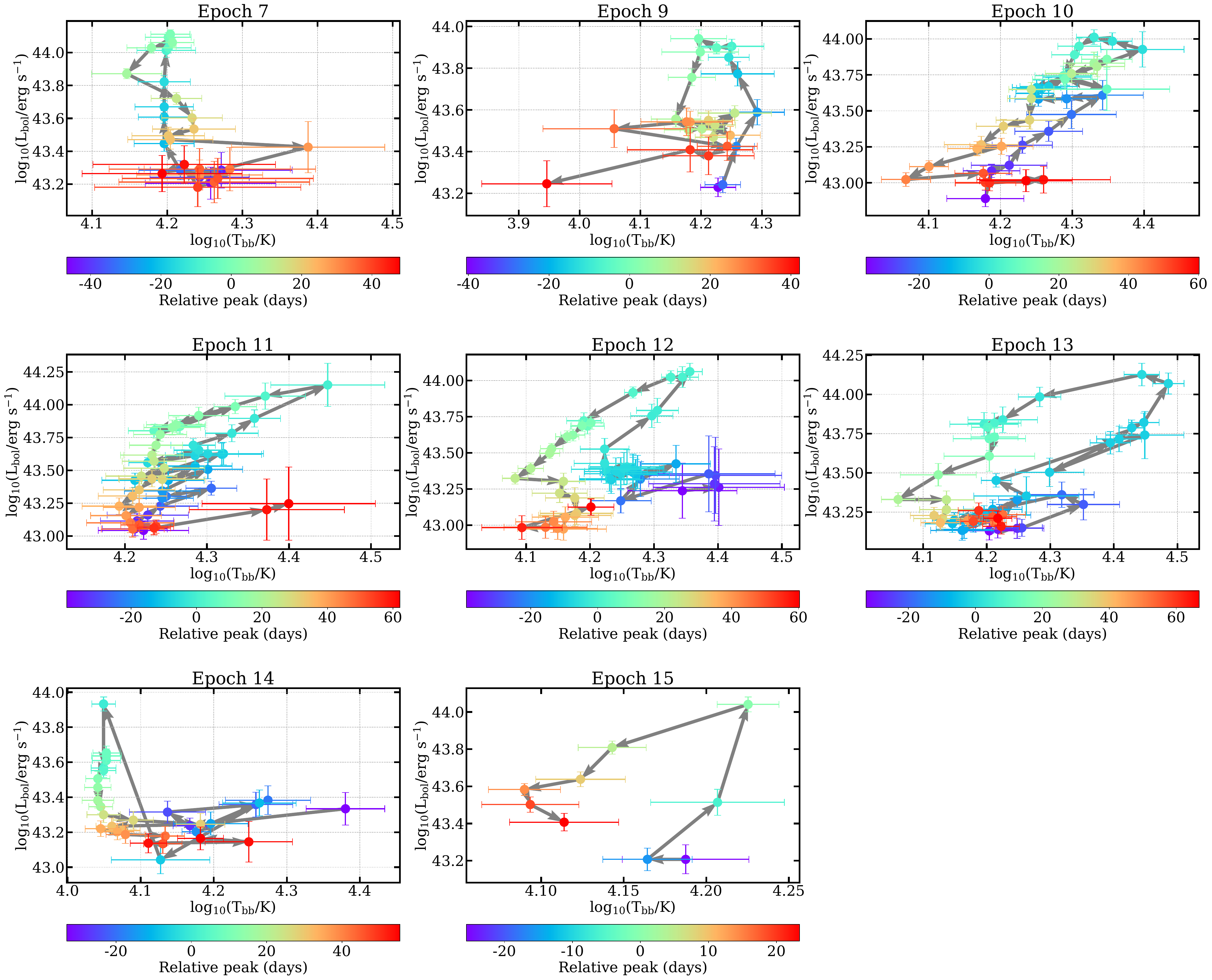}}
    \subfigure[]{
    \includegraphics[width=0.75\textwidth]{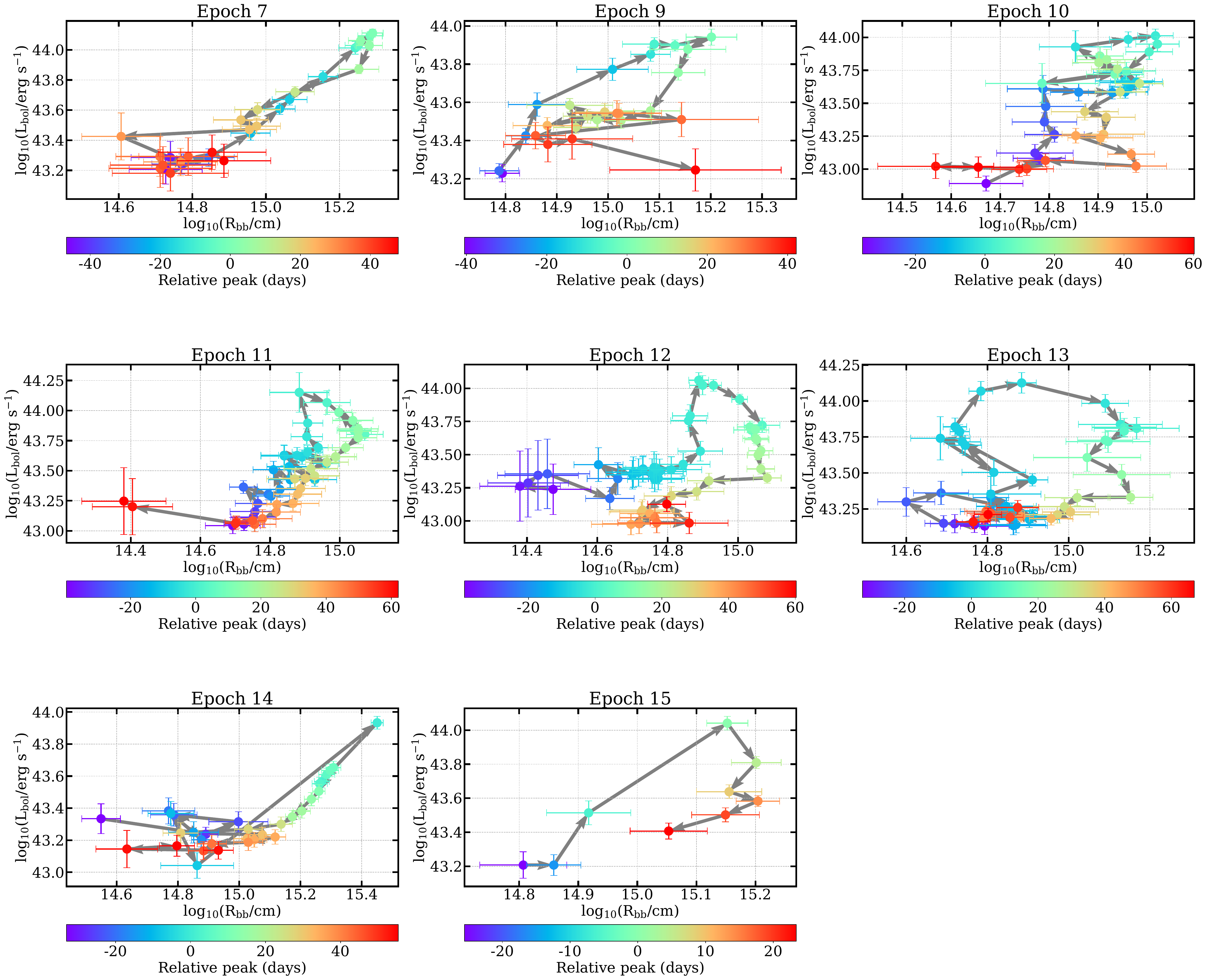}}
    \caption{(a): Evolution of blackbody luminosity with temperature and time in different epochs. (b): Evolution of blackbody luminosity with blackbody radius and time in different epochs.}
    \label{fig:lbb_tbb}
\end{figure*}

High-cadence multiwavelength observations provide us with an excellent opportunity to track the details of the temporal evolution of the UV/optical emission region of ASASSN-14ko. In Figure \ref{fig:lbb_tbb}, we plot the evolution of $L_{\rm bb}$, $T_{\rm bb}$ and $R_{\rm bb}$. In epochs 9--13, a significant counterclockwise evolution tendency that $L_{\rm bb}$ increases with $T_{\rm bb}$ and $R_{\rm bb}$, then decreases with $T_{\rm bb}$ and $R_{\rm bb}$. For instance, in epoch 10, $T_{\rm bb}$ is $10^{4.18\pm0.05}~\rm K$ in MJD 59926.88 with $L_{\rm bb}=10^{42.89\pm{0.06}}~\rm erg~s^{-1}$ and $R_{\rm bb}=10^{14.67\pm{0.08}}~\rm cm$, and then increases to $10^{4.4\pm{0.06}}~\rm K$ in MJD 59959.43 with $L_{\rm bb}=10^{43.93\pm{0.12}}~\rm erg~s^{-1}$ and $R_{\rm bb}=10^{14.85\pm{0.07}}~\rm cm$. After reaching the maximum value, $T_{\rm bb}$ decreases, but $L_{\rm bb}$ does not. The luminosity continues to increase to a value of $10^{44.01\pm{0.05}}~\rm erg~s^{-1}$ with $T_{\rm bb}=10^{4.33\pm{0.02}}~\rm K$ and $R_{\rm bb}=10^{15.02\pm{0.04}}~\rm cm$ in MJD 59962.58. After that, both $T_{\rm bb}$ and $L_{\rm bb}$ decrease and finally in MJD 60025.3, we derive $L_{\rm bb}=10^{43.01\pm{0.08}}~\rm erg~s^{-1}$, $T_{\rm bb}=10^{4.24\pm{0.06}}~\rm K$ and $R_{\rm bb}=10^{14.66\pm{0.09}}~\rm cm$. We should note that except for epoch 11, maximum $L_{\rm bb}$ did not occur in the maximum $T_{\rm bb}$ and $R_{\rm bb}$, which is similar to QPEs, especially for the case of eRO-QPE3 and eRO-QPE4 \citep{Arcodia2024}. The comparison with QPEs is seen in Figure~\ref{fig:kt_lx}. 

We estimated the expansion velocity of the photosphere using the black-body radius during the bumps. The mean expansion velocities for epochs 7, and 9-13 are $0.027c$, $0.012c$, $0.009c$, $0.003c$, $0.004c$, and $0.007c$, respectively. During epochs 9-13, for the rebrightenings, we derived the photosphere velocities as $-0.001c$, $-0.0008c$, $-0.005c$, $0.002c$, and $-0.006c$.

\subsection{Sporadic outbursts in X-rays}
\begin{figure}
    \centering
    \includegraphics[width=0.47\textwidth]{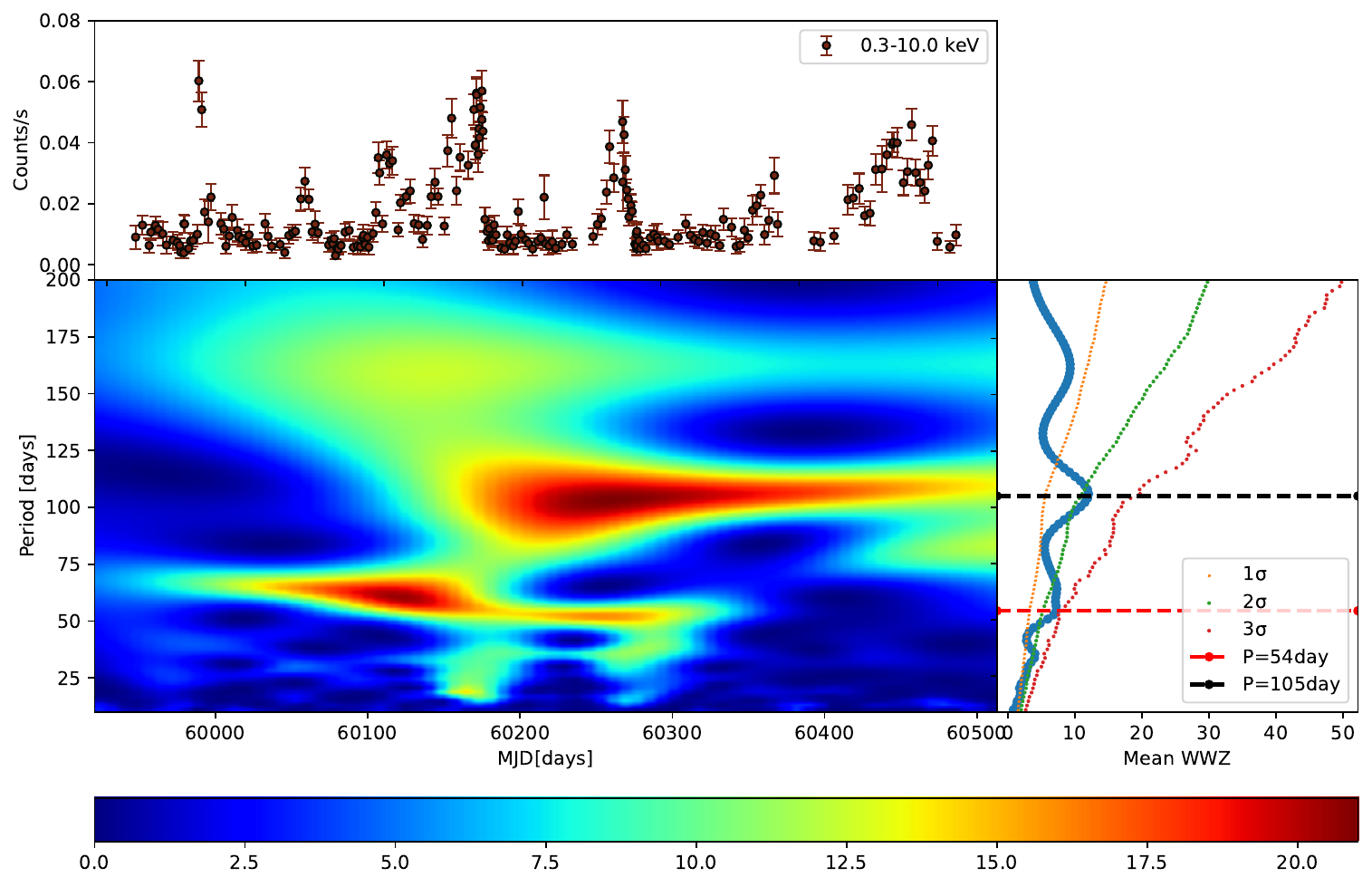}
    \caption{WWZ results of X-ray light curve.}
    \label{fig:xray_wwz}
\end{figure}

\cite{Payne2021} initially reported a period of approximately 114 days in the optical bands, which was later refined to around 115 days with additional outburst data \citep{Payne2023}. Their studies revealed that X-rays exhibit different behavior compared to UV/optical bands. Through high-cadence multiwavelength observations, \cite{Huang2023} noticed a potential quasi-periodicity of roughly two months in the X-ray. We employed the weighted wavelet Z-transform (WWZ) method \citep{Foster1996} with a Python package \texttt{ wwz}\footnote{\url{https://github.com/skiehl/wwz}} \citep{Kiehlmann2023} to search for QPO signals in the X-ray light curves. We used the Monte Carlo (MC) method to estimate the significance of the QPOs in the WWZ map, with detailed steps provided in \cite{Chen2024}. The WWZ periodogram revealed two QPO signals at 54 days and 105 days, both with significance levels exceeding $2\sigma$. The results can be seen in Figure \ref{fig:xray_wwz}.

\subsection{Relation Between UV/optical Bands and X-rays}
\begin{figure*}
\centering
\includegraphics[width=0.8\textwidth]{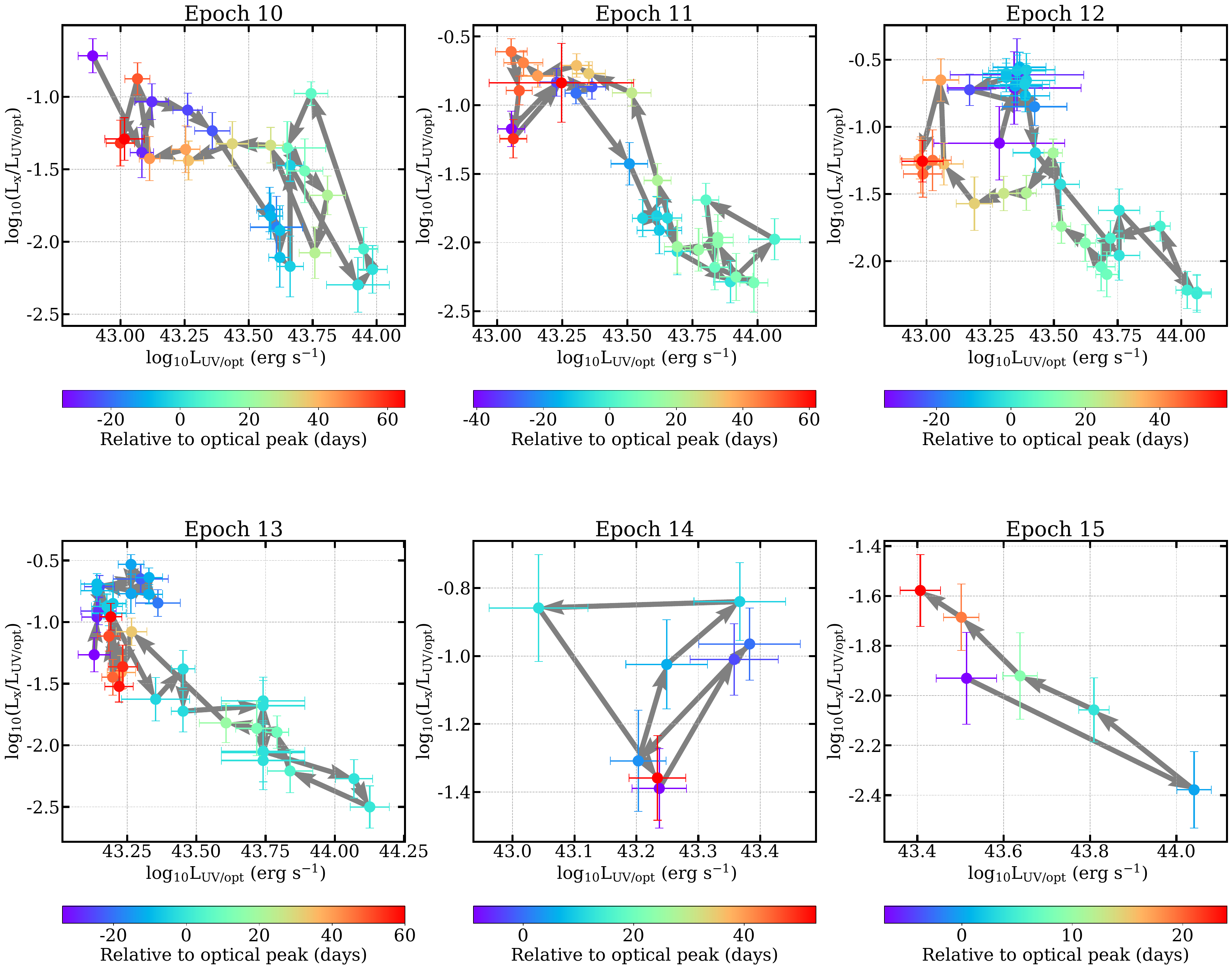}
\caption{Evolution of $L_{\text{x}}/L_{\text{UV/opt}}$ in different epochs.
In epochs 10--13, high-cadence, and continued
multiwavelength observations were performed. Epochs 14 and 15 lack enriched multiwavelength data but nevertheless show a similar overall evolution tendency.}
\label{fig:lx_lopt}
\end{figure*}

The overall UV/optical light curves show a pattern quite different from the X-rays. When a peak occurs in UV/optical bands, the X-rays exhibit a rather low state (see Figure \ref{fig:phaes}). It may indicate some potential correlation between UV/optical and X-rays. Therefore, we adopted the luminosity ratio of X-rays to the UV/optical bands ($L_{\text{x}}/L_{\text{UV/opt}}$) to investigate the multiwavelength behavior. The results in epochs 10--13 can be seen in Figure \ref{fig:lx_lopt}. $L_{\text{x}}/L_{\text{UV/opt}}$ exhibited a tendency that decreases as the UV / optical bands increase, as reported in \cite{Huang2023}. In each epoch, the value of $L_{\text{x}}/L_{\text{UV/opt}}$ initially decreases as $L_{\text{UV/opt}}$ increases and then increases as $L_{\text{UV/opt}}$ decreases. In epochs 10--13, the value of $\log_{10}(L_{\text{x}}/L_{\text{UV/opt}})$ ranges from -2.5 to -0.5, and through high-cadence observations, the temporal evolution is shown.  

\section{Discussion}\label{sec:discussion}
Observations of the early evolution of TDEs are crucial for understanding the physical mechanisms underlying these transient accretion processes. However, because of their unpredictability, the majority of TDEs lack early multiwavelength observations. However, ASASSN-14ko offers an exceptional opportunity to study the complete evolution of a TDE. Through high-cadence, multiwavelength observations, we have obtained the complete light curve of ASASSN-14ko across multiple outbursts. Our findings indicate that the bumps and rebrightenings in the light curve are diminishing, consistent with predictions of stream debris impacting the accretion disk in the rpTDE. In contrast, the X-ray light curves are sporadic, with outbursts which are offset from the UV/optical ones by intervals which vary from flare to flare.
Our high-cadence, multiwavelength observations have enabled continuous monitoring of the rpTDE, revealing the luminosity evolution trend with temperature and black hole radius, akin to QPEs. In this section, we discuss the physical origins of these phenomena.

\subsection{Physical Process Behind the Repeated Bumps and Rebrightenings}\label{4.1}
In the context of rpTDE, \cite{Huang2023} suggested that stream debris impacts the expanded disk and can produce repeated bumps and rebrightenings in UV/optical bands. Each encounter of the star within the tidal radius results in the tidal stripping of stellar material by the BH. Over multiple rpTDEs, this stripped material gradually coalesces around the BH, causing the accretion disk to expand. Once the disk size surpasses a critical threshold, the incoming fallback stream debris makes contact with the disk's outer edge, producing an early bump prior to the main UV/optical outburst, which is driven by the accretion process. Moreover, as the trailing stream follows the orbit of the leading stream, its impact on the disk can give rise to rebrightening in the UV/optical outburst. 

Each collision between the stream and the disk, analogous to the events observed in OJ 287, generates a ``bubble" allowing UV/optical photons to escape \citep{Lehto1996,Valtonen2008,Valtonen2016,Dey2018}. However, the ``bubble's" expansion leads to a decrease in luminosity corresponding to a drop in temperature (Figure \ref{fig:lbb_tbb}). The energy released during each collision is given by $\Delta E_{\rm s} \approx \Delta M_{\rm s} v_{\rm rel}^2/2$, where $\Delta M_{\rm s}$ is the mass of the shocked material,  $v_{\rm rel}$ is the stream-disk velocity difference \citep{Huang2023}. It is important to note that the velocity of the stream significantly influences the released energy. 

As depicted in Figure \ref{fig:intervals}, there is an overall decreasing trend in $E_{\rm bump}$ in several recent outbursts, potentially linked to changes in the orbital velocity of the stream debris. The precession of the stream's orbit, a consequence of general relativity, causes it to strike the disk at varying locations and velocities. The disk's own precession further diversifies the impact sites with each crossing. The Lense-Thirring effect causes the stellar orbital plane to undergo nodal precession, which influences the stream-disk interaction by altering both the impact site and the angle. At different radii from the black hole, the debris stream collides at different velocities; hence, as the impact site moves farther out, the collision velocity decreases and the resulting bumps and rebrightenings diminish. Moreover, relativistic precession alters the impact angle, which in turn modifies the collision velocity and thus the energy released in these bumps and rebrightenings. Thus, the diminishing $E_{\rm bump}$ could suggest increasingly distant impact locations on the disk. However, the complex environment of the disk results in after the initial impact of the stream, the velocity of the trailing stream, and consequently the energy of the subsequent rebrightening $E_{\rm reb}$ may not follow a uniform trend with $E_{\rm bump}$. Moreover, the stream's orbital precession also alters the timing of early bumps and rebrightenings. If this model holds, we anticipate a more pronounced pattern of bumps and rebrightenings in future events, especially when the impact site moves closer to the BH once again.

After each rpTDE, the remnant of the disrupted star continues to orbit the BH, with collisions between the star and the disk occurring at every pass. We attempted to apply post-Newtonian formalism to estimate the orbital parameters, assuming that star-disk interactions cause repeated bumps and rebrightenings. In reality, the disrupted star follows an eccentric orbit \citep{Cufari2022, Liu2023}, and at least one of the impact sites should be very close to the pericenter. Within this case, the star encounters the disk twice within a time span shorter than that of a circular orbit. Consequently, either $t_{\rm bump}$ or $t_{\rm reb}$ should be very short (about few days). However, the observed durations of bumps and rebrightenings are longer than those predicted by theoretical calculations for such an eccentric orbit. Additionally, the star-disk interaction produces lower luminosity and energy than observed \citep{Huang2023}. Therefore, we rule out the possibility that the repeated bumps and rebrightenings are caused by collisions between the disrupted star and the disk.

\subsection{Orbit of Stream Debris}\label{4.2}
Following the above scenario, we can derive the orbital parameters of the stream debris by analyzing the timing of the bumps and rebrightenings. According to \cite{Dey2018}, the time delay of the flare after the collision is significantly influenced by both the mass of the impacting body and its velocity. Given that the stripped mass is negligible relative to the BH mass, it is reasonable to assume that the emissions resulting from the stream-disk collision are promptly released. By recording the times of impact during each orbit, and considering a BH mass of $10^{7.86}~M_\sun$ \citep{Payne2021} and an orbital period of 115.2 days \citep{Payne2023}, we can constrain the orbital parameters of the system.

In our scenario, the stripped material is composed of front and trailing streams. The specific orbital energy of the streams is distributed in the range of ($E_{*}-\Delta E$, $E_{*}+\Delta E$), where $E_{*}=-GM_{BH}/2a$ is the specific orbital energy of the star and $\Delta E=GM_{BH}R_{*}/R_{p}^{2}$ is the characteristic energy surplus of the two streams. Here, $G$ is the gravitational constant, $ M_{BH}$ is the BH mass, $a$ is the semi-major axis of the orbit, $R_{*}$ is the stellar radius and $R_{\rm p}$ is the pericenter radius. Therefore, the duration of the fallback of two streams is 
\begin{equation}
\Delta t_{\rm fb}=\frac{3}{2}\times \frac{2\Delta E}{|E_{*}|}P=3\left(\frac{a}{R_{\rm p}}\right)\left(\frac{R_{\rm p}}{a}\right)_{\rm crit}P,
\end{equation}
where $P$ is the orbital period and the subscript ``crit" represents the value when $\Delta E=|E_{*}|$. For the case of ASASSN-14ko, we derive $R_{\rm p}/a\approx 0.02$ and $(R_{\rm p}/a)_{\rm crit}=2.2\times 10^{-3}(m_{*}/M_8)^{1/3}$, where $m_*$ is the stellar mass in unit $1~M_\sun$ and $M_8$ is the BH mass in unit $10^8M_\sun$. The time interval between bump and rebrightening in every UV/optical outburst refers to $\Delta t_{\rm fb}$, which is approximately 1/3 of the stellar orbital period. The calculation is consistent with the results of our observations (Figure \ref{fig:intervals}).

Consider the case of a spinning BH where both the disk and stellar orbits are misaligned with the direction of the BH spin. Therefore, the nodal precession of the stellar orbit caused by Lense-Thirring effect produces an angle of 
\begin{equation}
\Delta \Omega \simeq \frac{\pi}{2} a_{*}\left(\frac{R_S}{R_p}\right)^{3/2},
\end{equation}
where $a_{*}$ is the dimensionless spin of the BH and $R_S$ is the Schwarzchild radius. When the disk is extended enough, the occurrence of bumps and rebrightenings are periodic and it would back to the previous level after the number of stellar orbits of 
\begin{equation}
N=\frac{2\pi}{\Delta\Omega} \simeq 40 \left(\frac{0.8}{a_{*}}\right)\left(\frac{R_p}{4R_S}\right)^{3/2}.
\end{equation}

\subsection{The Origin of Sporadic  Behaviour in the X-Ray }\label{4.3}
The periodicity analysis of the X-ray light curve, covering MJD 59900--60513, reveals periodicities at 54 days and 105 days. The latter aligns closely with the 111.6-day optical period identified by \cite{Payne2021} using the Lomb-Scargle Periodogram (LSP). Additionally, \cite{Huang2023} reported a two-month periodicity, also discerned through an extensive LSP analysis. In contrast with the UV/optical bands, the X-ray emissions exhibit sporadic behavior rather than the predictable outburst pattern observed in the former. It should be noted that the intervals between peak X-ray emissions are markedly irregular, and after epoch 12, there is a pronounced elongation in the duration between X-ray outbursts.

In the multiwavelength light curves, the main UV/optical outbursts exhibit a trend inverse to that of the X-rays. Specifically, X-ray outbursts were detected in the quiescent phases of the UV/optical bands, and conversely, when UV/optical outbursts peak occur, the X-ray emission subsides to its quiescent state. This dichotomous behavior is clearly illustrated in Figure \ref{fig:lx_lopt} in each epoch. Moreover, our high-cadence X-ray observations, as depicted in Figure \ref{fig:lc_mw}, reveal a ``softer-when-brighter" pattern, a trend also observed and reported by \cite{Payne2023}. 

The anomalous X-ray activity may be attributable to disk and/or coronal structural changes triggered by a TDE \citep{Ricci2020,Li2022,Masterson2022,Cao2023}. The (partial) disruption of a star by the central
BH may lead to the destruction of the inner accretion disk, causing the overlying corona to dissipate. This would result in a reduction of non-thermal X-ray emission. Subsequently, as the inner disk reforms, thermal flux would increase while the corona is being reconstructed, returning the X-ray emissions to their high, non-thermal state \citep{Ricci2020,Cao2023}. Given that the host galaxy of ASASSN-14ko is an AGN with detectable hard X-ray emissions indicative of an
existing corona, the aforementioned scenario may be applicable to this source
\citep{Payne2021}. Nevertheless, we note that for ASASSN-14ko, the recovery time for the disk and corona reconstruction in this scenario, of just a few days, would be much faster than that observed in a comparable source \citep{Ricci2020}.

\subsection{Evolution of the Blackbody Temperature}\label{4.4}
Historically, the erratic nature of TDEs has led to a lack of comprehensive observation, limiting our understanding of these extreme events. However, the high-cadence multiwavelength observations of ASASSN-14ko represent a breakthrough, providing us with a unique opportunity to better investigate the entire process of TDEs.

A distinct feature of TDEs is the constant  blackbody temperature in the UV/optical bands, which can distinguish them from supernovas and AGNs \citep{Velzen2020,Zabludoff2021}. In Figure \ref{fig:lbb_tbb}, the blackbody luminosity of ASASSN-14ko evolves with temperature, which shows some differences from previous typical cases and more resembles QPEs \citep{Miniutti2019,Arcodia2024}. However, the blackbody temperature of ASASSN-14ko is still rather high with $>10^4~\rm K$ without substantial deviations beyond an order of magnitude, aligning with the optical selected TDEs previously reported. The evolution of temperature is also accompanied by the evolution of blackbody radius, which indicates the variation of radiation region, although at a mild level. In past 6 epochs with high-cadence multiwavelength observations, we can clearly record the evolution of luminosity with the photosphere, and such rpTDE provides an excellent opportunity to investigate the hydrodynamic process and radiation mechanism of the extreme accretion event.

\section{Conclusion}\label{sec:conclusion}
ASASSN-14ko is a periodic nuclear transient observable in both UV/optical and X-ray bands. Its most prominent feature is the approximately 115-day periodicity in UV/optical light curves, a phenomenon that has been extensively investigated in several studies. This periodicity is attributed to a rpTDE, offering a unique opportunity to study the physical processes involved in TDEs.

Through long-term, high-cadence multiwavelength observations, we have detected multiple repeated bumps and rebrightenings in the UV/optical light curves. The energy released in bumps and rebrightenings exhibits an overall decrease tendency. The occurrence time of bumps and rebrightenings is variable and does not show a monotonous trend. These phenomena are unprecedented in previous outbursts and other known TDEs. This unique variation in the UV/optical light curves is likely caused by stream debris interacting with the expanded disk in a rpTDE. The rpTDE leads to disk expansion, where the front stream collides with the disk, producing the early bump, and the trailing stream subsequently collides with the disk, causing the rebrightening.

We analyzed the X-ray light curve and detected periodicities of 54 and 105 days (only $>2\sigma$ confidence level) using the WWZ method. The X-ray luminosity exhibits an opposite trend compared to the UV/optical bands. 
Consequently, this process may produce outflows, which could also explain the observed radio emission (Section~\ref{4.3}).

Another significant discovery from our high-cadence multiwavelength observations is the evolution of the blackbody temperature and radius. During each UV/optical outburst, the luminosity increases with the blackbody temperature, and a similar behavior is observed in the blackbody radius. This evolution of luminosity is not seen in previously reported TDEs but is a common phenomenon in QPEs.

As a rpTDE, ASASSN-14ko offers an excellent opportunity to study the physical processes and radiation mechanisms of TDEs. \cite{Steinberg2024} found that the existence of the disk increases the efficiency of the debris circularization and then reduces the rising time in the TDE. In previous UV/optical outbursts, the short rising time scale (within 10 days) is much shorter than typical TDEs \citep{Velzen2019,Huang2023b,Yao2023}. The simulation result of \cite{Steinberg2024} is consistent with the observation in ASASSN-14ko. Additionally, its periodicity allows for multiwavelength, high-frequency observations in advance of the predicted outburst periods, which is crucial for investigating the early evolution of TDEs. The periodicity and evolution of ASASSN-14ko closely resemble those of QPEs, suggesting that this source may represent a special type of QPE, characterized by periodic outbursts in both X-ray and UV/optical bands. Long-term multiwavelength monitoring is essential to fully understand the physical properties of such sources.

\section{Acknowledgements}
We thank the anonymous referee for providing valuable comments and suggestions, which helped to improve the manuscript. The authors thank Dr. Norbert Schartel for his contribution on the Chandra data. We also thank Prof. Jiliang Jing for his comments and suggestions on this work. The authors thank Brad Cenko and the Swift team for approving our proposal and executing the observations. This work is supported by the National SKA program of China (2022SKA0130102), the Strategic Priority Research Program of the Chinese Academy of Sciences (XDB0550200), the National Key Research and Development Program of China (2023YFA1608100), the National Natural Science Foundation of China (grants 12393814, 12192221, 12103048, 12073091, 12375046), Anhui Provincial Natural Science Foundation (2308085QA32), Shandong Provincial Natural Science Foundation
(ZR2023MA036), and the Fundamental Research Funds for Central Universities (WK2030000097). We gratefully acknowledge the support of Cyrus Chun Ying Tang Foundations, the Frontier Scientific Research Program of the Deep Space Exploration Laboratory (2022-QYKYJH-HXYF-012).
The authors acknowledge the use of public data from the Swift data archive.  This work has made use of data from the Asteroid Terrestrial-impact Last Alert System (ATLAS) project. The Asteroid Terrestrial-impact Last Alert System (ATLAS) project is primarily funded to search for near earth asteroids through NASA grants NN12AR55G, 80NSSC18K0284, and 80NSSC18K1575; by products of the NEO search include images and catalogs from the survey area. This work was partially funded by Kepler/K2 grant J1944/80NSSC19K0112 and HST GO-15889, and STFC grants ST/T000198/1 and ST/S006109/1. The ATLAS science products have been made possible through the contributions of the University of Hawaii Institute for Astronomy, the Queen’s University Belfast, the Space Telescope Science Institute, the South African Astronomical Observatory, and The Millennium Institute of Astrophysics (MAS), Chile.
This scientific work uses data obtained from Inyarrimanha Ilgari Bundara / the Murchison Radio-astronomy Observatory. We acknowledge the Wajarri Yamaji People as the Traditional Owners and native title holders of the Observatory site. CSIRO’s ASKAP radio telescope is part of the Australia Telescope National Facility (\url{https://ror.org/05qajvd42}). Operation of ASKAP is funded by the Australian Government with support from the National Collaborative Research Infrastructure Strategy. ASKAP uses the resources of the Pawsey Supercomputing Research Centre. Establishment of ASKAP, Inyarrimanha Ilgari Bundara, the CSIRO Murchison Radio-astronomy Observatory and the Pawsey Supercomputing Research Centre are initiatives of the Australian Government, with support from the Government of Western Australia and the Science and Industry Endowment Fund. 
This paper includes archived data obtained through the CSIRO ASKAP Science Data Archive, CASDA (\url{http://data.csiro.au}).
This research has made use of data obtained from the Chandra Data Archive provided by the Chandra X-ray Center (CXC). 

\vspace{5mm}
\facilities{Swift, Chandra, ATLAS, ASASSN, ASKAP}

\software{astropy \citep{2013A&A...558A..33A,AstropyII,AstropyIII}, HEASoft \citep{2014ascl.soft08004N}, Xspec \citep{1996ASPC..101...17A}, Matplotlib \citep{2007CSE.....9...90H}.
          }

\appendix

\section{Blackbody Luminosity}
\begin{figure*}
    \centering
    \figurenum{A1}
    \subfigure[]{
    \includegraphics[width=0.3\linewidth]{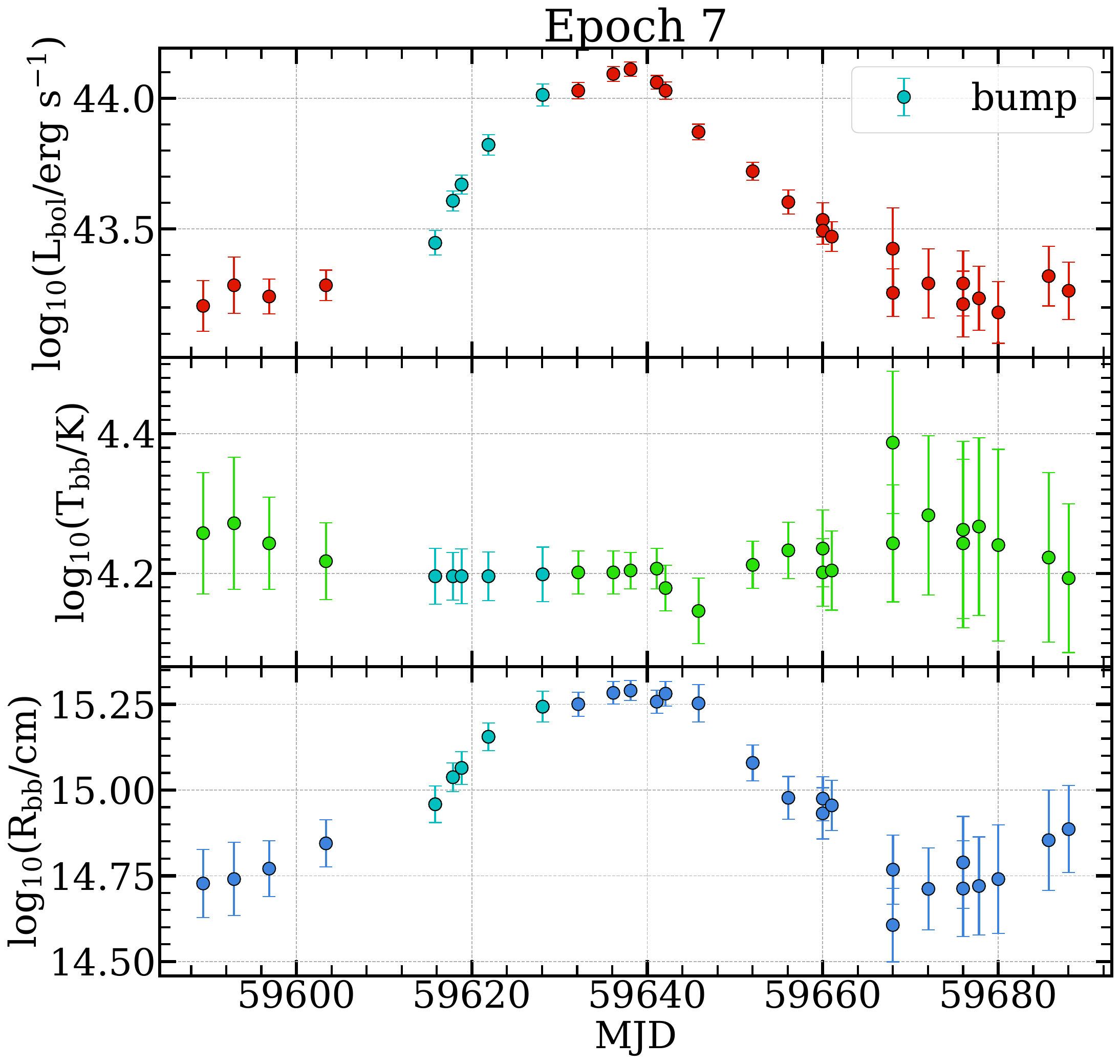}}
     \subfigure[]{
    \includegraphics[width=0.3\textwidth]{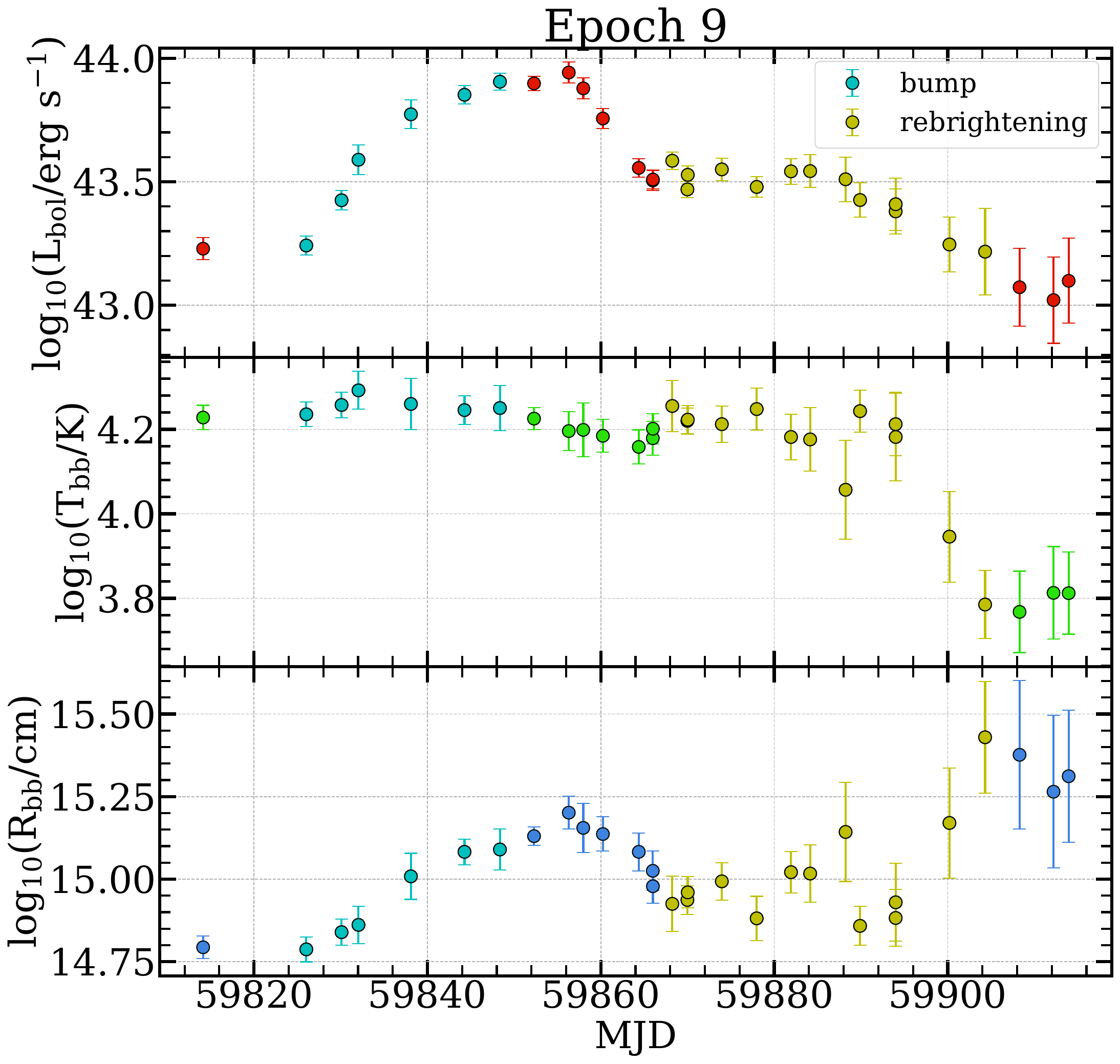}}
     \subfigure[]{
    \includegraphics[width=0.3\textwidth]{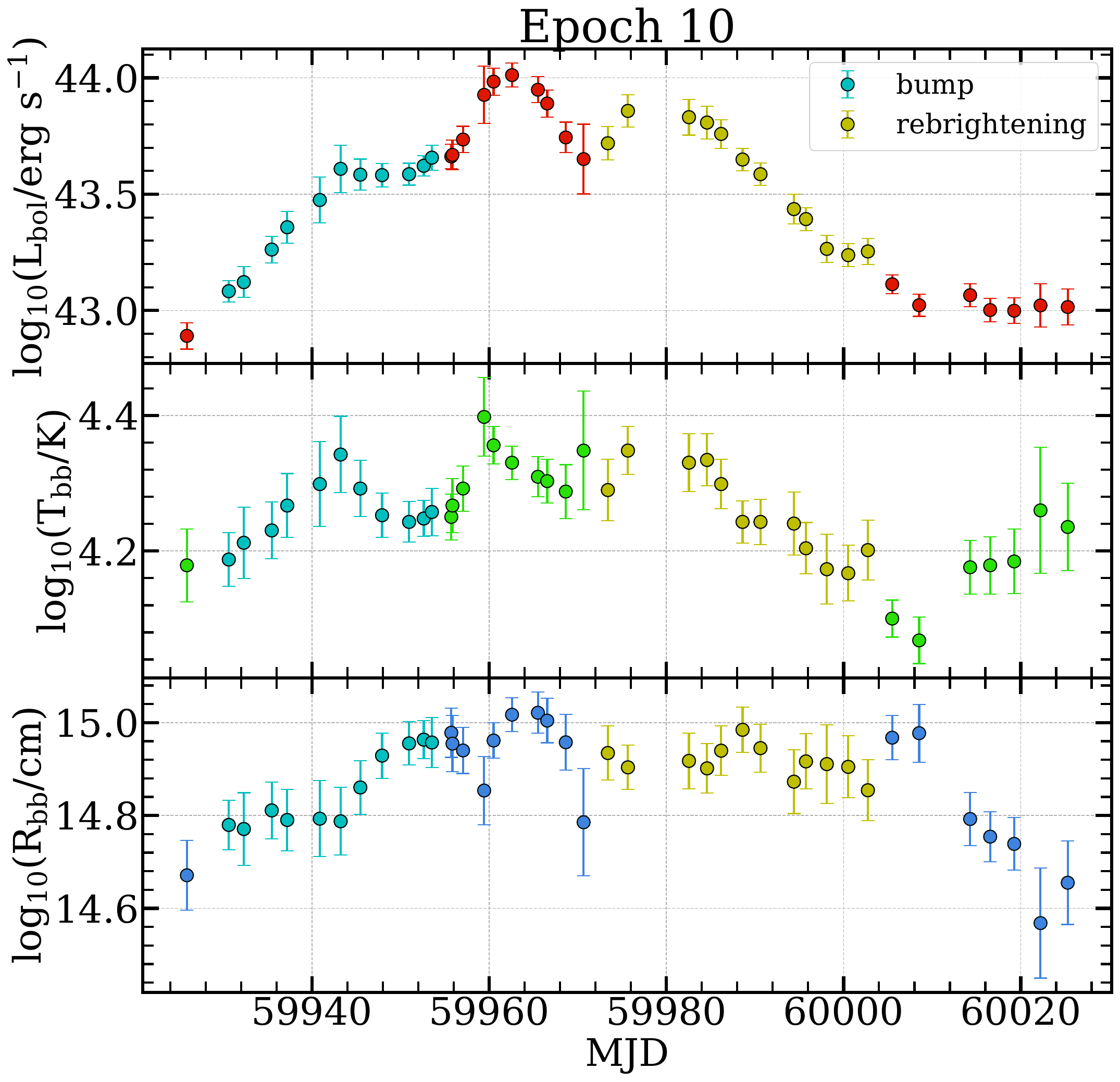}}
    
     \subfigure[]{
    \includegraphics[width=0.3\textwidth]{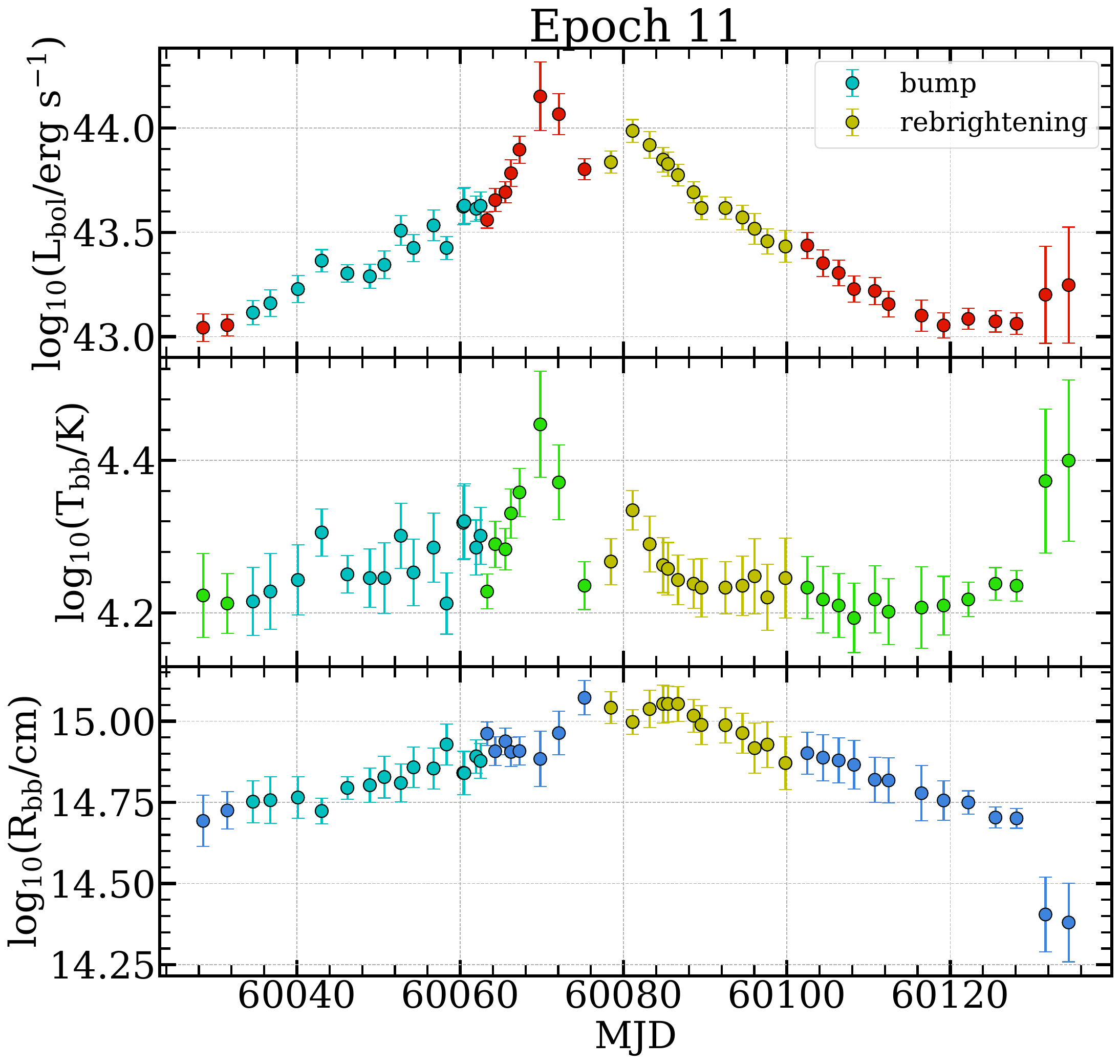}}
     \subfigure[]{
    \includegraphics[width=0.31\textwidth]{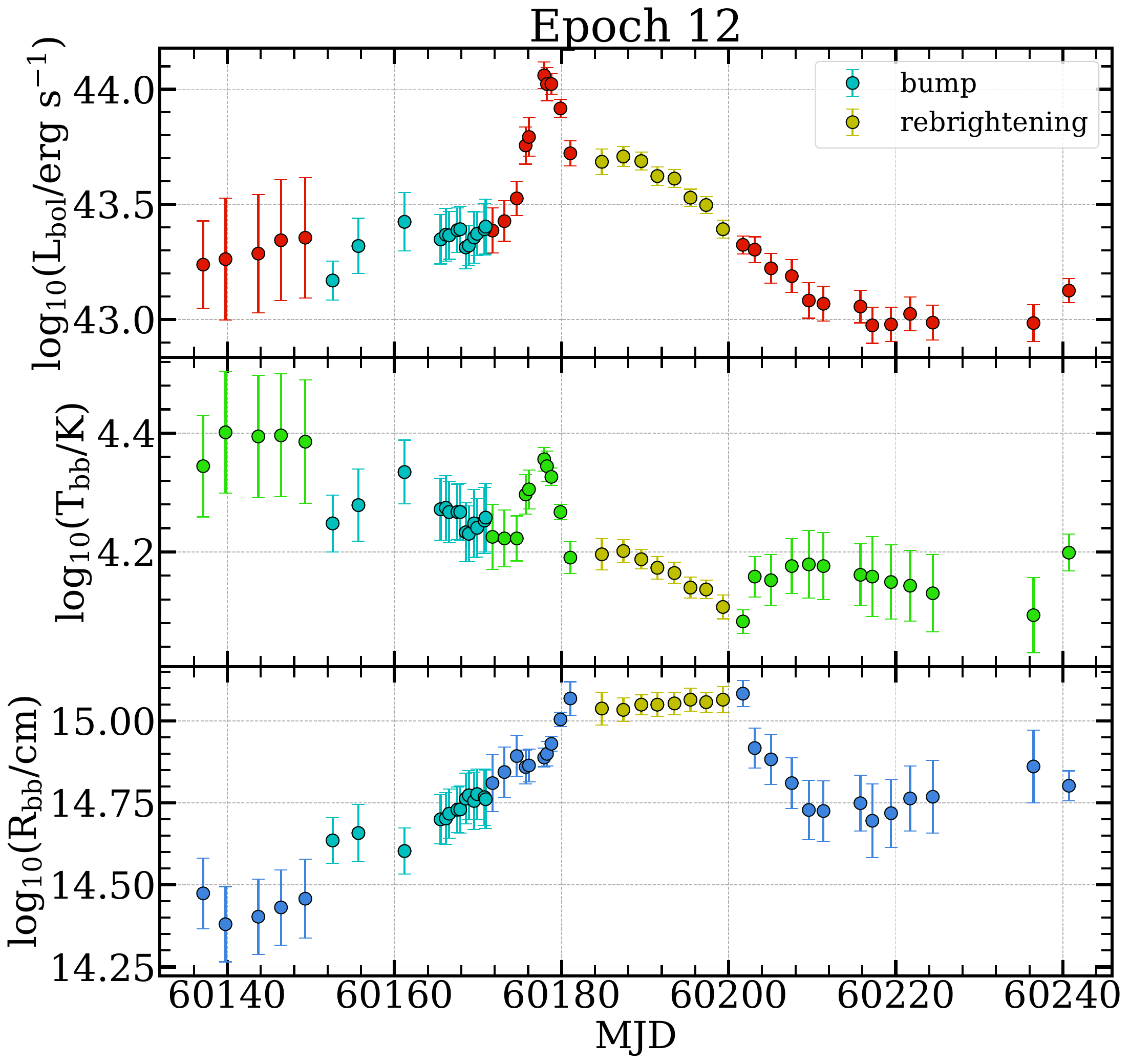}}
     \subfigure[]{
    \includegraphics[width=0.315\textwidth]{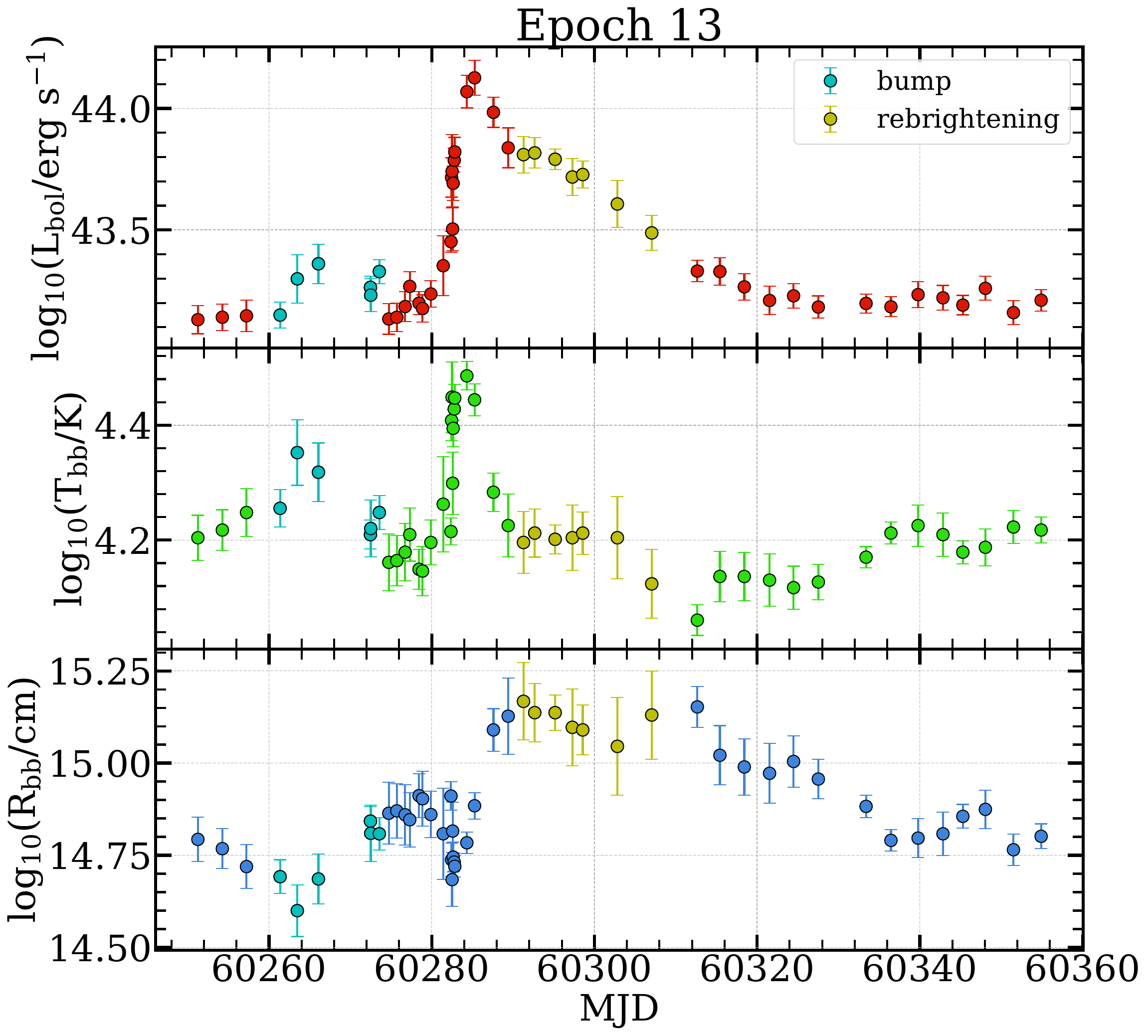}}
    \caption{In each subfigure, from the top panel to the bottom plot the UV/optical black body luminosity, temperature and black body radius of ASASSN-14ko. Data points in the period of bump and rebrightening are marked by different colors.}
    \label{fig:lc_bb}
\end{figure*}

Covering from MJD 58957.46 to 60384, the host-subtracted data in \textsl{o}, \textsl{c}, \textsl{g}, \textsl{B}, \textsl{U}, \textsl{UVW1}, \textsl{UVM2}, and \textsl{UVW2} bands were fitted by \texttt{SuperBol} \citep{Nicholl2018}.  Due to a significant gap in epoch 14, which caused the missing detection of the main UV/optical outburst by Swift, we did not fit the multiwavelength data in this epoch. 6 epochs with rich UV/optical data were fitted and the results are shown in Figure \ref{fig:lc_bb}.

\section{Comparison with QPEs}
\begin{figure*}
    \centering
    \figurenum{B1}
    \includegraphics[width=0.85\textwidth]{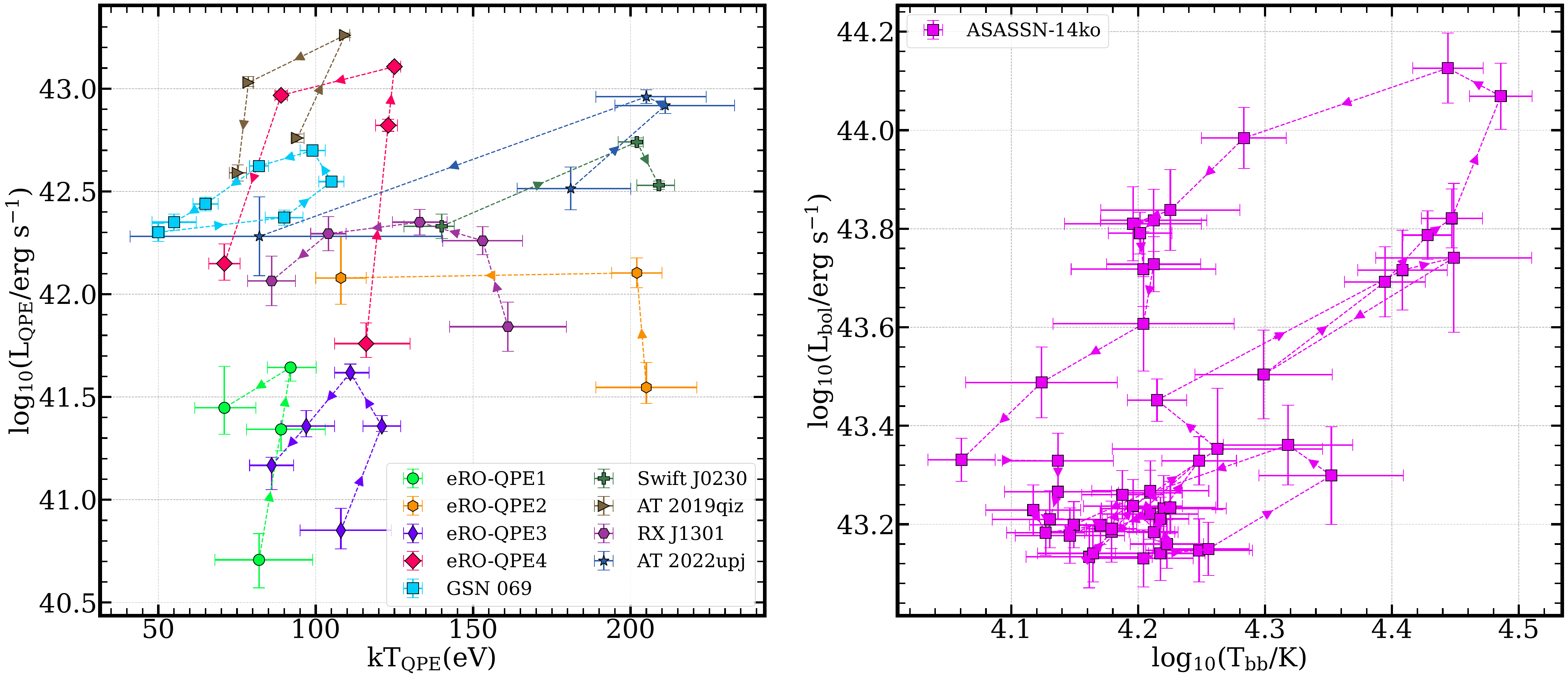}
    \caption{The evolution of temperature with the luminosity for part of reported QPEs and ASASSN-14ko. The data sources are as follows: eRO-QPE1 from \cite{Chakraborty2024}, eRO-QPE2 from \cite{Arcodia2024b}, eRO-QPE3 and eRO-QPE4 from \cite{Arcodia2024}, GSN~069 from \cite{Miniutti2019}, Swift~J0230 from \cite{Guolo2024}, AT~2019qiz from \cite{Nicholl2024}, RXJ1301 from \cite{Giustini2024} and AT~2022upj from \cite{Chakraborty2025}. The data for ASASSN-14ko is from epoch 13.}
    \label{fig:kt_lx}
\end{figure*}

The luminosity of ASASSN-14ko evolves with its temperature. In Figure~\ref{fig:kt_lx}, we compare ASASSN-14ko at epoch 13 with reported QPEs, including eRO-QPE1, eRO-QPE2 \citep{Arcodia2021}, eRO-QPE3, eRO-QPE4 \citep{Arcodia2024}, GSN~069 \citep{Shu2018,Miniutti2019}, RX~J1301.9+2747 \citep[hereafter RX~J1301;][]{Sun2013,Giustini2020}, AT~2019qiz \citep{Nicholl2024}, AT~2022upj \citep{Chakraborty2025}, and Swift~J023017.0+283603 \citep[hereafter Swift~J0230;][]{Evans2023,Guolo2024}. All QPE data sets were obtained from publicly available literature. Most QPEs in our sample exhibit an increase in luminosity with rising temperature and the trend is also observed in ASASSN-14ko. In the $kT_{\text{QPE}}-L_{\text{QPE}}$ (or $T_{\text{bb}}-L_{\text{bb}}$) diagrams, a counterclockwise evolution is evident in most QPEs and ASASSN-14ko, except for Swift~J0230.

\section{X-ray evolution during UV/optical outbursts}

\begin{figure*}
    \centering
    \figurenum{C1}
    \includegraphics[width=0.85\textwidth]{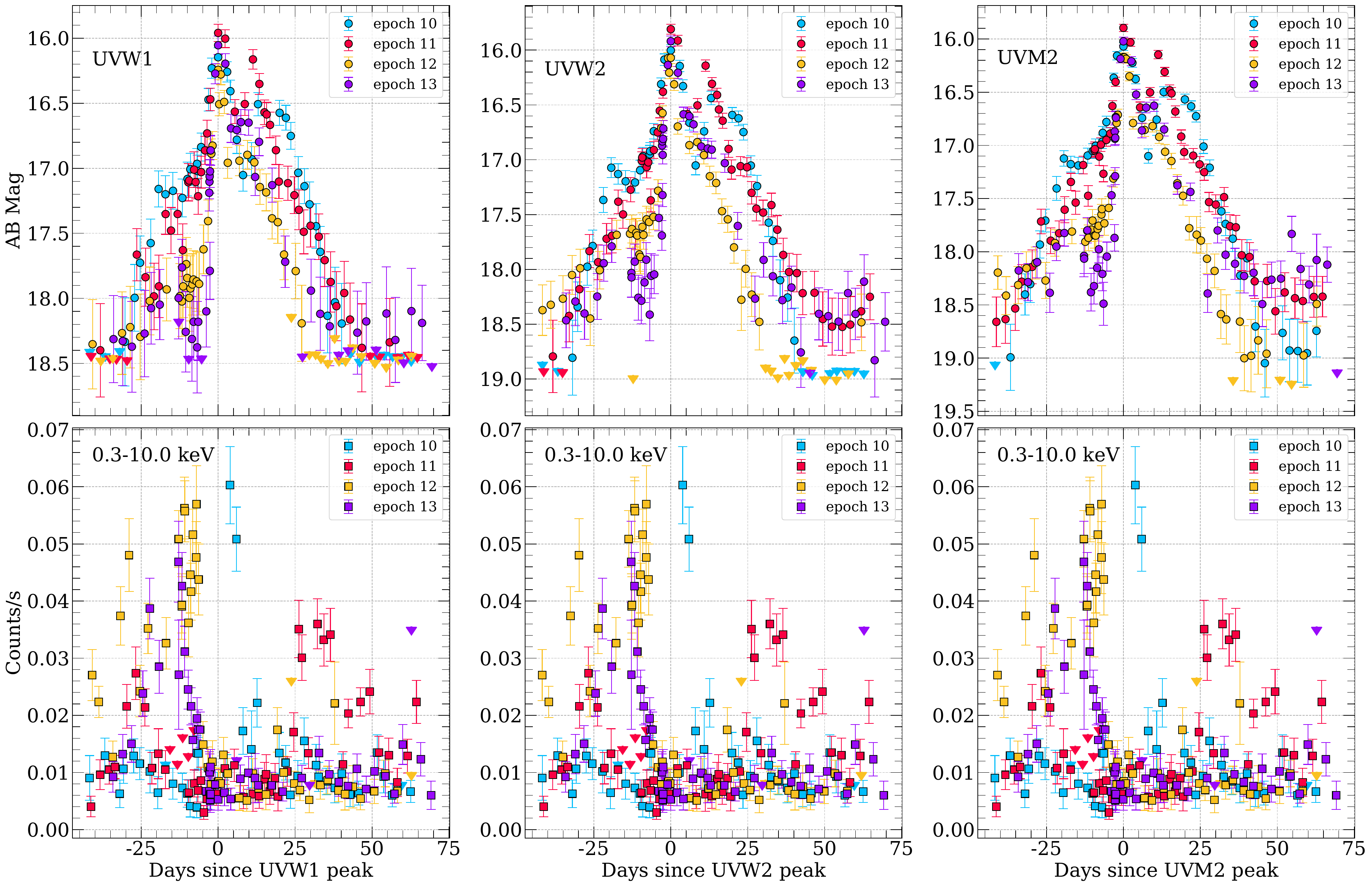}
    \caption{The comparison between the X-rays and UV bands during epoch 10--13.}
    \label{fig:phaes}
\end{figure*}

The multiwavelength light curves of ASASSN-14ko exhibit an anti-correlated behavior between X-ray variability and UV/optical bands. As an illustrative case, we present in Figure \ref{fig:phaes} the comparative evolution of UVW1, UVW2, and UVM2 bands with X-ray emission during epochs 10--13.  Except for epoch 10, significant X-ray flux enhancements consistently occur during the rising or declining phases of UV light curves, with no detectable significant variability in X-ray observed during UV peak phases.

\bibliography{ASASSN-14ko}{}
\bibliographystyle{aasjournal}

\end{document}